\documentclass[12pt,onecolumn]{IEEEtran}
\usepackage{makecell,amsbsy,amsmath,amssymb,epsfig,bbm,mathrsfs,multirow,amsthm,bm}
\usepackage{array,multirow,graphicx,MnSymbol}
\usepackage[english]{babel}
\usepackage[justification=centering]{caption}
\usepackage{url}
\usepackage{threeparttable}
\usepackage{epstopdf}
\usepackage{subfigure}
\usepackage[ruled,linesnumbered]{algorithm2e}
\usepackage{color,xcolor}
\usepackage{hyperref}

\DeclareMathAlphabet{\mathpzc}{OT1}{pzc}{m}{it}
\SetKwInOut{Parameter}{Parameter}
\SetKwInOut{Initialization}{Initialization}

\usepackage{url}

\IEEEoverridecommandlockouts

\begin{document}
\bibliographystyle{IEEE2}

\title{Achieving Covert Communication in Large-Scale SWIPT-Enabled D2D Networks}

\author{Shaohan Feng,~\IEEEmembership{Member,~IEEE}, Xiao Lu,~\IEEEmembership{Member,~IEEE}, Dusit Niyato,~\IEEEmembership{Fellow,~IEEE}, Ekram Hossain,~\IEEEmembership{Fellow,~IEEE}, and Sumei Sun,~\IEEEmembership{Fellow,~IEEE}}

\maketitle

\vspace{-15mm}
\begin{abstract}\vspace{-2mm}
We aim to secure a large-scale device-to-device (D2D) network against adversaries. The D2D network underlays a downlink cellular network to reuse the cellular spectrum and is enabled for simultaneous wireless information and power transfer (SWIPT). In the D2D network, the transmitters communicate with the receivers, and the receivers extract information and energy from their received radio-frequency (RF) signals. In the meantime, the adversaries aim to detect the D2D transmission. The D2D network applies power control and leverages the cellular signal to achieve covert communication  (i.e., hide the presence of transmissions) so as to defend against the adversaries. We model the interaction between the D2D network and adversaries by using a two-stage Stackelberg game. Therein, the adversaries are the followers minimizing their detection errors at the lower stage and the D2D network is the leader maximizing its network utility constrained by the communication covertness and power outage at the upper stage. Both power splitting (PS)-based and time switch (TS)-based SWIPT schemes are explored. We characterize the spatial configuration of the large-scale D2D network, adversaries, and cellular network by stochastic geometry. We analyze the adversary's detection error minimization problem and adopt the Rosenbrock method to solve it, where the obtained solution is the best response from the lower stage. Taking into account the best response from the lower stage, we develop a bi-level algorithm to solve the D2D network's constrained network utility maximization problem and obtain the Stackelberg equilibrium. We present numerical results to reveal interesting insights. For example, the PS-based SWIPT scheme outperforms the TS-based SWIPT scheme in terms of both network performance (e.g., link reliability and power outage probability) and resistance to the adversary, i.e., steady network utility against increasing aggressiveness of the adversary. 
\end{abstract}


%\clearpage

\section{Introduction}
\label{sec:introduction}

The unprecedented popularity of mobile devices such as laptops, smart phones, and sensors is driving the need to explore future wireless networks to support wireless communications without any fixed infrastructure. D2D network has attracted significant research attention for its capability of enabling connection among proximal mobile devices in the absence of pre-existing infrastructure~\cite{feng2022securing}. The D2D network can work as a supplement to the cellular network and provision numerous novel Internet-based applications and services including but not limited to emergency calls and personal networking~\cite{SWIPT2023feng}. However, the D2D network faces the following challenges due to the on-device battery of these mobile devices~\cite{6609136}. First, the energy consumption of massive mobile devices is strongly adding to global warming and, moreover, the exhausted on-device battery makes an escalating contribution to environmental pollution. Second, due to the limited capacity of the on-device battery, it is required to frequently replace/recharge the on-device battery so as to ensure network sustainability, which, however, is impractical especially in uninhabited and dangerous areas~\cite{8896400}. This motivates us to explore a more energy-efficient and environmental-friendly solution for the energy supply in the D2D network. 

\subsection{SWIPT-Enabled D2D Networks}

To cope with the aforementioned challenges, we resort to SWIPT~\cite{9385384}, which is a promising approach to prolong the lifetime of mobile devices. The SWIPT enables a mobile device to extract information and energy from its received RF signal. In particular, the mobile device will divide its received signal into two parts, where one part is for information decoding and the other part is for energy harvesting. In this way, the energy supply of a mobile device can be guaranteed even without installing a large-capacity on-device battery such that the environmental pollution can be alleviated. Moreover, by enabling energy harvesting from the received signal, it is not necessary to frequently replace/recharge mobile devices. Motivated by these advantages, we study a SWIPT-enabled D2D network in this paper, which, however, faces security challenges such as eavesdropping and is difficult to be secured due to the following reasons: 
\begin{enumerate}
\item As the mobile devices in a SWIPT-enabled D2D network are wirelessly powered and hence are energy-constrained, it is difficult to perform cryptographic methods to achieve confidentiality on the data due to the high energy consumption incurred by the encryption and decryption. In addition, the effectiveness of the encryption will be compromised when the adversary has powerful computational capacities. 

\item Due to the decentralized architecture of a D2D network (i.e., without fixed infrastructure), it is complicated and impractical to implement the network security functionalities such as authentication and distribution and management of cryptographic key. This together with the broadcast nature of wireless medium imply open access of the data to the adversaries.

\item What is worse in the SWIPT-enabled D2D network is that the receiver strongly depends on its corresponding transmitter due to the dedicated power transfer. In this case, if we cannot hide the presence of wireless communication so as to avoid attacks such as locating the transmitter, the receiver will run out of power and finally fail.
% once it is physically attacked.
\end{enumerate}

\subsection{Covert Communication}

To tackle the challenges in securing the SWIPT-enabled D2D network, we resort to covert communication~\cite{9108996}, which is also referred to as undetectable communication and has drawn significant research attention in recent years. The covert communication safeguards the wireless networks by preventing the adversary from detecting legitimate communication while maintaining a certain transmission rate for the intended receiver. It offers  the following advantages. First, the covert communication promises stronger protection compared with the encryption method by hiding the presence of the legitimate communication, and the information transmitted is therefore immune from interception. As such, the security performance is independent of the information processing capacity of the adversaries. Moreover, the covert communication does not incur high energy consumption and hence is suitable for energy-constrained networks. Second, as covert communication does not rely on the implementation of fixed infrastructure and complex functionalities, it is applicable to decentralized networks. Third, covert communication can shelter not only the presence of the legitimate communication but also the position and movement of the legitimate transmitter. As such, both the transmitter and its dedicated receiver in the SWIPT-enabled D2D network can be secured. 

\subsection{Contributions}
\label{subsec:contribution}

Motivated by the aforementioned advantages, in this paper, we adopt covert communication to secure the SWIPT-enabled D2D network from transmission detection by adversaries~\cite{9385384}. In particular, the D2D network is large-scale, containing massive transmitters and receivers, and underlays a downlink cellular network with a large number of base stations (BSs) and cellular users (CUs) as shown in Fig.~\ref{fig:system_model}(a). Therein, the D2D transmitters and the BSs communicate with the D2D receivers and the CUs, respectively. The receivers receive the ambient signal from the environment, including the D2D network and cellular network, for jointly harvesting energy and decoding information. In the meantime, the adversaries aim to detect D2D transmission based on their received signal powers. The D2D network performs power control to achieve covert communication so as to hide the presence of the D2D transmissions. Moreover, the D2D network leverages the signal transmitted by the cellular network, namely cellular signal, to increase the interference dynamics at the adversaries so as to distort their observations and thereby mislead their decision-making, which will further enhance communication covertness. The major contributions of this paper are summarized as follows:
\begin{itemize}
\item We secure the D2D network via covert communication. By doing this, not only the data can be completely secured but also the presence of the D2D transmission can be hidden from the adversaries, which significantly enhances network security. 

\item We model the network spatial configuration by using stochastic geometry and derive the network performance measures accordingly so as to conduct the study from a system-level perspective.

\item We explore both PS-based and TS-based SWIPT schemes for the D2D network and demonstrate the advantages and disadvantages of these two schemes in terms of network security and sustainability.

\item We model the combat between the D2D network and adversaries in the framework of a two-stage Stackelberg game. The adversaries are the followers aiming to detect the D2D transmission at the lower stage while the D2D network is the leader aiming to maximize its network utility, considering reliability and power cost, subject to the constraints on power outage and communication covertness at the upper stage. 

\item We analyze the game equilibrium and develop a bi-level algorithm based on genetic algorithm (GA) and Rosenbrock method to search for the equilibrium. We verify the optimality of the obtained equilibrium. We also present numerical results to evaluate the network performance and reveal practical insights. 
\end{itemize}

\subsection{Organization and Notations}

The notations along with their descriptions and typical values are shown in Table~\ref{tab:notation_value}. Section~\ref{sec:related_work} presents the related works. The system model is illustrated in Section~\ref{sec:system_model}, which is followed by the problem formulation and algorithm design in Section~\ref{sec:game_equilibrium}. Section~\ref{sec:numerical_results} presents the numerical results  and Section~\ref{sec:conclusion} draws the conclusion and summarizes the insights.

%\clearpage

\section{Related Works}
\label{sec:related_work}

\subsection{SWIPT}
\label{subsec:related_SWIPT}

By enabling the simultaneous transfer of the information and power, SWIPT becomes a promising solution to support the energy-constrained networking and therefore has been applied in many application scenarios~\cite{6951347, 7457656, 7063588}. In the meantime, the enhanced methods for the SWIPT technology have been investigated widely~\cite{7457656, 7081080}. For example, in~\cite{7457656}, SWIPT has been applied to wirelessly power the D2D network and 3D sectorized antennas have been incorporated to address the challenges in the transfer efficiency of information and power incurred by the 3D network spatial configuration. Moreover, a 3D antenna sectorization method, including horizontal and vertical spatial separations, is developed to control the 3D sectorized antennas so as to further enhance the transfer efficiency. In~\cite{7063588}, SWIPT has been applied to multiple-input multiple-output (MIMO) broadcasting networks for simultaneously serving an energy-harvesting user and an information-decoding user over the same time-frequency channel. This work proposed a weighted minimum mean squared error (WMMSE) criterion aiming to minimize the weighted sum-mean squared error (MSE) for the information-decoding user while lower bounding the power outage for the energy-harvesting user. Smart antenna technologies, including MIMO and relaying, have been applied to improve the transfer efficiency of information and power for SWIPT~\cite{7081080}. The MIMO technologies are used to improve the concentration of the signal propagation so as to enhance the transfer efficiency of both the information and power. The relaying technologies enable a relay node to harvest energy from the RF signal from the source and then use the harvested energy to forward the information to the destination, which not only improves the efficiency of exploiting RF signal but also motivates the cooperation among the network nodes. 

\subsection{Covert Communication}
\label{subsec:related_CC}

Covert communication has attracted massive attention  as a new security paradigm for next-generation wireless networks. It has been applied widely~\cite{8355734, 9382022, 9361424, feng2022securing, 9736993, 9580594} and its enhanced methods have been extensively studied~\cite{7805182, 9361424, feng2022securing, 9736993, 9108996}. For example, in~\cite{8355734}, a one-way relay network is investigated, where a relay node is deployed to amplify and thereafter forward the source's message to the destination. The relay node is greedy and opportunistically transmits its own message to the destination via covert communication so as to shelter its illegitimate usage of the legitimate resource. In~\cite{7805182}, a new performance metric for measuring the communication covertness is defined, where the detection error incurred by the noise uncertainty is taken into account so as to evaluate the overall communication covertness. The authors in~\cite{9382022} study a UAV-assisted air-ground network, where the air-ground communication link is vulnerable to malicious users due to its high line-of-sight (LoS) probability. Covert communication has been applied to secure the air-ground communication link by hiding its presence. A scenario that a legitimate transmitter wants to covertly communicate with multiple receivers is considered in~\cite{9361424}, and a friendly jammer is deployed to enhance the communication covertness by exploiting the interference dynamics incurred by the jamming signal to distort the adversary's observation. A similar idea is adopted in~\cite{9736993} that the co-channel interference incurred by the concurrent transmission together with the artificial noise (AN) are leveraged to secure a large-scale IoT network, where the network spatial configuration is modeled by stochastic geometry. Covert communication has been applied to secure the federated learning process in~\cite{9580594}. As the covert communication needs to limit the transmission power so as to maintain the communication covertness while federated learning requiring frequent and up-to-date model update (i.e., reliable and low latency communication), an optimization problem is formulated to address these two contradictory targets. Due to the capability of achieving constructive (destructive) effect on the intended (unintended) users, an intelligent reflecting surface (IRS) is introduced in~\cite{9108996} to improve communication covertness. Therein, the phase shifts of the IRS elements are adjusted to concentrate the signal propagation over the intended user while weakening the signal leakage over the adversary. In this way, the legitimate communication can be improved while the detection performance of the adversary can be compromised. In~\cite{feng2022securing}, covert communication is used as a complementary security approach to the information-theoretical secrecy approach\footnote{Information-theoretical secrecy approach achieves a positive secrecy rate (i.e., a positive rate difference between the legitimate and wiretap channels).}. Specifically, it is used to secure D2D communication with the assistance from friendly jammers by sheltering the presence of the D2D communication as the primary level of protection. Moreover, an information-theoretical secrecy approach is applied to ensure the secrecy outage once the covert communication fails and the D2D communication has been detected, which is regarded as the secondary level of protection. 

Inspired by these works, we adopt the SWIPT technique and covert communication to prolong the lifetime of and secure the large-scale D2D network, respectively. This is the major motivation and the key contribution of this work.

%\clearpage 

\section{System Model and Assumptions}
\label{sec:system_model}

\subsection{Network Description}

\begin{figure}[!]
	\centering
	\includegraphics[width=1\textwidth,trim=90 40 100 40, clip]{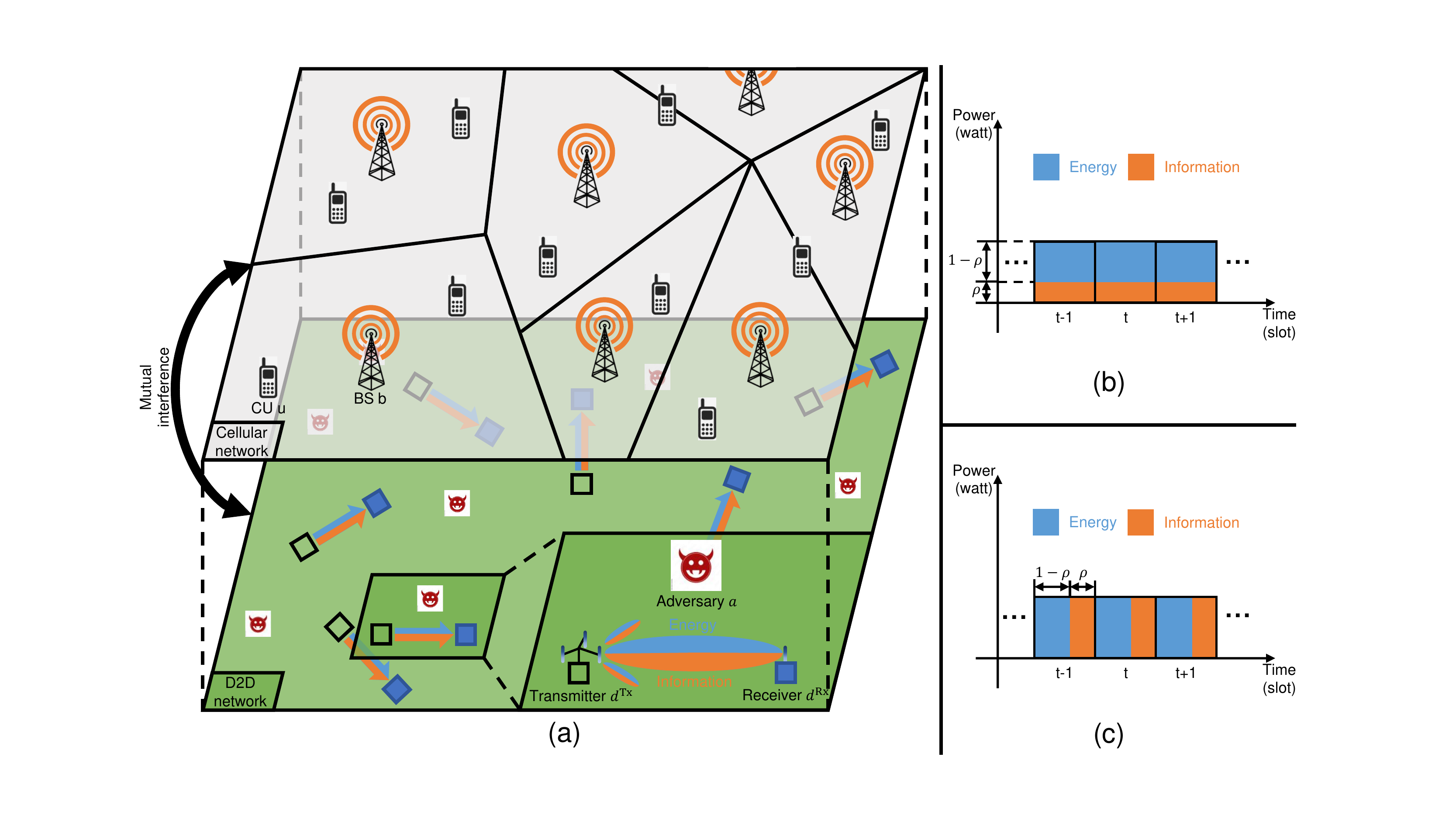}
	\caption{(a) SWIPT-enabled D2D network underlaying cellular network threatened by adversaries, (b) PS-based SWIPT scheme, and (c) TS-based SWIPT scheme.}
	\label{fig:system_model}
\end{figure}

We aim to secure a bi-dimensional wireless D2D network as shown in Fig.~\ref{fig:system_model}(a), which underlays a downlink cellular network~\cite{9385384} and is threatened by the adversaries. The D2D and cellular networks as well as the adversaries operate over the same time-frequency resource block (RB). Therein, the D2D network contains a massive number of transmitters  the set of which is denoted as ${\cal{D}}^{\rm{Tx}}$. The locations of the transmitters are modeled as a homogeneous Poisson point process (HPPP) $\Phi_{{\cal{D}}^{\rm{Tx}}}$ with density $\lambda_{{\cal{D}}^{\rm{Tx}}}$. Each transmitter serves a dedicated receiver at distance $R$ in an arbitrary orientation~\cite{7457656}. The transmitters randomly become active and communicate with their dedicated receivers, and the events that the transmitter is active and inactive are denoted by ${\cal{H}}_1$ and ${\cal{H}}_0$ and occur with the probabilities of ${\mathbb{P}}^{{\cal{H}}_1}$ and ${\mathbb{P}}^{{\cal{H}}_0} = 1 - {\mathbb{P}}^{{\cal{H}}_1}$, respectively. The adversaries following another independent HPPP $\Phi_{\cal{A}}$ with density $\lambda_{\cal{A}}$ are passive and attempt to detect the D2D transmissions from the transmitters to their dedicated receivers. For the cellular network, the BSs follow an independent HPPP $\Phi_{\cal{B}}$ with density $\lambda_{\cal{B}}$ and the CUs follow another independent HPPP $\Phi_{\cal{U}}$ with density $\lambda_{\cal{U}}$~\cite{SWIPT2023feng}, where ${\cal{B}}$ and ${\cal{U}}$ are the sets of BSs and CUs, respectively. Moreover, we consider a nearest-BS association policy and $\lambda_{\cal{U}} \gg \lambda_{\cal{B}}$ such that each BS has at least one CU in its Voronoi cell to serve~\cite{SWIPT2023feng}. Similarly, each BS independently becomes active, denoted by ${\cal{C}}_1$, and inactive, denoted by ${\cal{C}}_0$, over the considered RB with probabilities ${\mathbb{P}}^{{\cal{C}}_1}$ and ${\mathbb{P}}^{{\cal{C}}_0} = 1 - {\mathbb{P}}^{{\cal{C}}_1}$, respectively. To analyze the network performance, we condition on that the receiver at the origin is the representative receiver, namely the typical receiver, and denoted by receiver $d^{\rm{Rx}}$. Moreover, the transmitter associated by receiver $d^{\rm{Rx}}$ is regarded as the typical transmitter and denoted by transmitter $d^{\rm{Tx}}$. For practicality, we consider that transmitter $d^{\rm{Tx}}$ aims to avoid transmission detection against its nearest (i.e., most threatening) adversary, which is thereby denoted by adversary $a$. On the other hand, we consider that adversary $a$ attempts to detect the transmission of transmitter $d^{\rm{Tx}}$. Time is partitioned into slots of duration $\Delta t$ with $T$ and $t$ denoting the number of slots and slot index, respectively (i.e., $t\in\left\{1, \ldots, T\right\}$). Similar to the works related to the system performance analysis~\cite{4712724}, the discussion in the rest of this paper is based on the performance of the representative network nodes (i.e., receiver $d^{\rm{Rx}}$, transmitter $d^{\rm{Tx}}$, and adversary $a$).

\subsection{Channel Model}

Each transmitter is equipped with multiple ($M$) antennas while the receivers, the BSs, and the CUs are equipped with a single omnidirectional antenna. For the signal from the D2D network, we use the received signal power at receiver $d^{\rm{Rx}}$ locating at ${\bf{x}}_{d^{\rm{Rx}}}$ regarding its associating transmitter $d^{\rm{Tx}}$ locating at ${\bf{x}}_{d^{\rm{Tx}}}$ as an example, and the rest can be defined similarly. In particular, the received signal power at receiver $d^{\rm{Rx}}$ with respect to (w.r.t.) transmitter $d^{\rm{Tx}}$ is $\left\|{\bf{h}}_{d^{\rm{Tx}}d^{\rm{Rx}}}^{\rm{H}} {\bf{w}}_{d^{\rm{Tx}}}\right\|^2 \ell\left({\bf{x}}_{d^{\rm{Tx}}}, {\bf{x}}_{d^{\rm{Rx}}}\right)$, where ${\bf{w}}_{d^{\rm{Tx}}} \in {\mathbb{C}}^{M \times 1}$ is the beamformer, ${\bf{h}}_{d^{\rm{Tx}}d^{\rm{Rx}}} \in {\mathbb{C}}^{M \times 1}$ is the small-scale fading, ${\rm{H}}$ is the conjugate transpose, and $\ell\left({\bf{x}}_{d^{\rm{Tx}}}, {\bf{x}}_{d^{\rm{Rx}}}\right) = \left\|{\bf{x}}_{d^{\rm{Tx}}} - {\bf{x}}_{d^{\rm{Rx}}}\right\|^{-\alpha} = R^{-\alpha}$ measures the large-scale fading with $\alpha$ being the path-loss exponent. Here, each entry of ${\bf{h}}_{d^{\rm{Tx}}d^{\rm{Rx}}}$ follows an independent and identical complex Gaussian distribution with zero mean and unit variance~\cite{7150338} (i.e., ${\bf{h}}_{d^{\rm{Tx}}d^{\rm{Rx}}} \sim {\cal{CN}}\left({\bf{0}}_M, {\bf{I}}_M\right)$ with ${\bf{0}}_M$ and ${\bf{I}}_M$ being zero vector and identity matrix of dimension $M$, respectively). For the signal from the cellular network, we use the received signal power at receiver $d^{\rm{Rx}}$ w.r.t. BS $b \in {\cal{B}}$ located at ${\bf{x}}_b$ as an example. Specifically, the received signal power at receiver $d^{\rm{Rx}}$ from BS $b$ is $p^{\rm{C}} h_{b d^{\rm{Rx}}}\ell\left({\bf{x}}_b, {\bf{x}}_{d^{\rm{Rx}}}\right)$, where $p^{\rm{C}}$ is the cellular transmission power, and $h_{b d^{\rm{Rx}}} \sim \exp\left(1\right)$ and $\ell\left({\bf{x}}_b, {\bf{x}}_{d^{\rm{Rx}}}\right) = \left\|{\bf{x}}_b - {\bf{x}}_{d^{\rm{Rx}}}\right\|^{-\alpha}$ measure the small-scale and large-scale fading between BS $b$ and receiver $d^{\rm{Rx}}$, respectively~\cite{SWIPT2023feng}.

\subsection{SWIPT Model}

The receivers not only receive the desired signal from their associating transmitters but also the interference from other transmitters as well as that from the BSs. To jointly enhance the power transfer efficiency and communication performance, the transmitters adopt the maximum ratio transmission (MRT) to perform SWIPT, e.g., the beamformer of transmitter $d^{\rm{Tx}}$ is ${\bf{w}}_{d^{\rm{Tx}}} = \sqrt{p^{\rm{S}}} {\bf{v}}_{d^{\rm{Tx}}}$, where $p^{\rm{S}}$ is the SWIPT transmission power and ${\bf{v}}_{d^{\rm{Tx}}} = \frac{{\bf{h}}_{d^{\rm{Tx}}d^{\rm{Rx}}}}{\left\|{\bf{h}}_{d^{\rm{Tx}}d^{\rm{Rx}}}\right\|} \in {\mathbb{C}}^{M \times 1}$ is the normalization of the channel between transmitter $d^{\rm{Tx}}$ and its serving receiver (i.e., receiver $d^{\rm{Rx}}$)~\cite{4712724}. Accordingly, the instantaneous received signal power at receiver $d^{\rm{Rx}}$ regarding the activation status of its corresponding transmitter is 
\begin{equation}
y_{d^{\rm{Rx}}} = \left\{
\begin{aligned}
& I^{\rm{S}}_{d^{\rm{Rx}}} + I^{\rm{C}}_{d^{\rm{Rx}}} + N^{\rm{RF}} + N_{d^{\rm{Rx}}}, & {\rm{if}} \, {\cal{H}}_0, \\
& p^{\rm{S}} g_{d^{\rm{Tx}}d^{\rm{Rx}}} \ell\left({\bf{x}}_{d^{\rm{Tx}}}, {\bf{x}}_{d^{\rm{Rx}}}\right) +  I^{\rm{S}}_{d^{\rm{Rx}}} + I^{\rm{C}}_{d^{\rm{Rx}}} + N^{\rm{RF}} + N_{d^{\rm{Rx}}}, & {\rm{if}} \, {\cal{H}}_1,
\end{aligned}\right.
\end{equation}
where $I^{\rm{S}}_{d^{\rm{Rx}}} = p^{\rm{S}} \sum\limits_{{d^{\rm{Tx}}}' \in \left\{\left.{\cal{D}}^{\rm{Tx}}\backslash \left\{d^{\rm{Tx}}\right\}\right| d^{\rm{Rx}} \right\}} {\mathbbm{1}}_{{d^{\rm{Tx}}}'} g_{{d^{\rm{Tx}}}'d^{\rm{Rx}}} \ell\left({\bf{x}}_{{d^{\rm{Tx}}}'}, {\bf{x}}_{d^{\rm{Rx}}}\right)$ is the interference from other transmitters, $I^{\rm{C}}_{d^{\rm{Rx}}} = p^{\rm{C}} \sum\limits_{b \in \left\{\left.{\cal{B}}\right| d^{\rm{Rx}} \right\}} {\mathbbm{1}}_b h_{b d^{\rm{Rx}}} \ell\left({\bf{x}}_b, {\bf{x}}_{d^{\rm{Rx}}}\right)$ is the interference from the BSs, $N^{\rm{RF}}$ is the additional noise in the RF energy harvesting circuit, and $N_{d^{\rm{Rx}}}$ is the additive noise. Therein, $\left\{{\bf{x}}_{{d^{\rm{Tx}}}'}\right\}_{{d^{\rm{Tx}}}' \in \left\{\left.{\cal{D}}^{\rm{Tx}}\backslash \left\{d^{\rm{Tx}}\right\}\right| d^{\rm{Rx}} \right\}}$ is one of the realizations of the PP of transmitters that excludes transmitter $d^{\rm{Tx}}$ with receiver $d^{\rm{Rx}}$ as the observation point (i.e., $\Phi_{\left\{\left.{\cal{D}}^{\rm{Tx}}\backslash \left\{d^{\rm{Tx}}\right\}\right| d^{\rm{Rx}} \right\}}$) and $\left\{{\bf{x}}_b\right\}_{b \in \left\{\left.{\cal{B}}\right| d^{\rm{Rx}} \right\}}$ is one of the realizations of the PP of BSs with receiver $d^{\rm{Rx}}$ as the observation point (i.e., $\Phi_{\left\{\left.{\cal{B}}\right| d^{\rm{Rx}} \right\}}$). ${\mathbbm{1}}_{{d^{\rm{Tx}}}'}$ is the activation indicator of transmitter ${d^{\rm{Tx}}}'$ and equals $1$ if it is active and $0$ otherwise, and ${\mathbbm{1}}_{b}$ can be defined similarly for BS $b$. Moreover, $g_{d^{\rm{Tx}}d^{\rm{Rx}}} = \left\|{\bf{h}}_{d^{\rm{Tx}}d^{\rm{Rx}}}^{\rm{H}} {\bf{v}}_{d^{\rm{Tx}}}\right\|^2 = \left\|{\bf{h}}_{d^{\rm{Tx}}d^{\rm{Rx}}}\right\|^2$ follows a Gamma distribution with shape parameter $M$\footnote{$M$ is the number of antennas at the transmitters.} and scale parameter $1$, and $g_{{d^{\rm{Tx}}}'d^{\rm{Rx}}} = \left\|{\bf{h}}_{{d^{\rm{Tx}}}'d^{\rm{Rx}}}^{\rm{H}} {\bf{v}}_{{d^{\rm{Tx}}}'}\right\|^2$ is exponentially distributed for ${d^{\rm{Tx}}}' \neq d^{\rm{Tx}}$~\cite{875282}.

The transmitters have dedicated power supplies, while the receivers are battery-free with rechargeable abilities (i.e., no energy storage for future use and all the energy harvested in a time slot will be consumed within this time slot~\cite{7320989}). Here, we consider two SWIPT schemes (i.e., PS-based and TS-based SWIPT schemes as shown in Figs.~\ref{fig:system_model}(b) and (c), respectively). 

\subsubsection{PS-Based SWIPT Scheme} 

\begin{figure}
     \centering
     \begin{minipage}{8cm}
		\centering
		\includegraphics[width=1\textwidth,trim=10 25 20 5,clip]{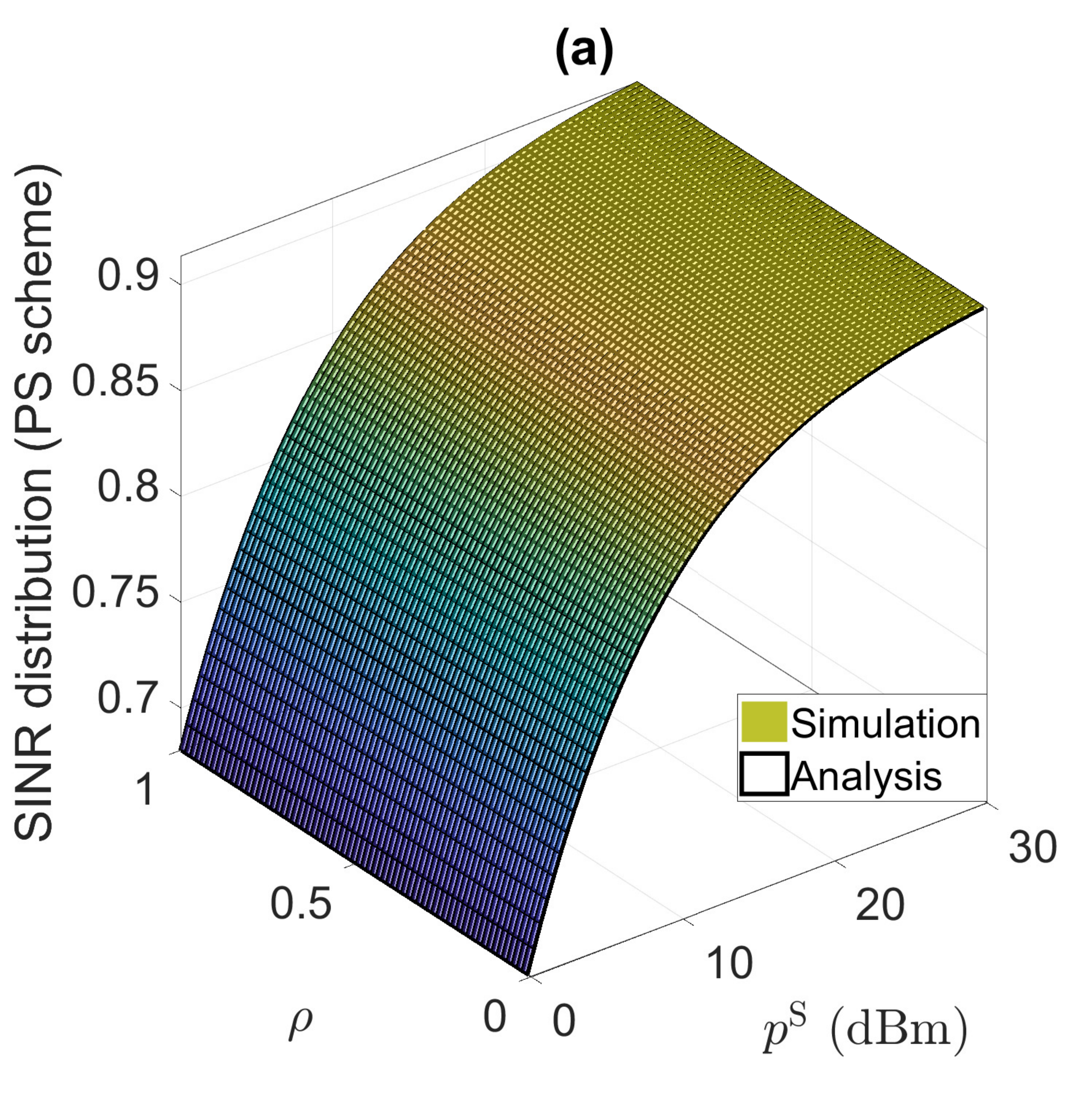}
     \end{minipage}
     \begin{minipage}{8cm}
		\centering
		\includegraphics[width=1\textwidth,trim=10 25 20 5,clip]{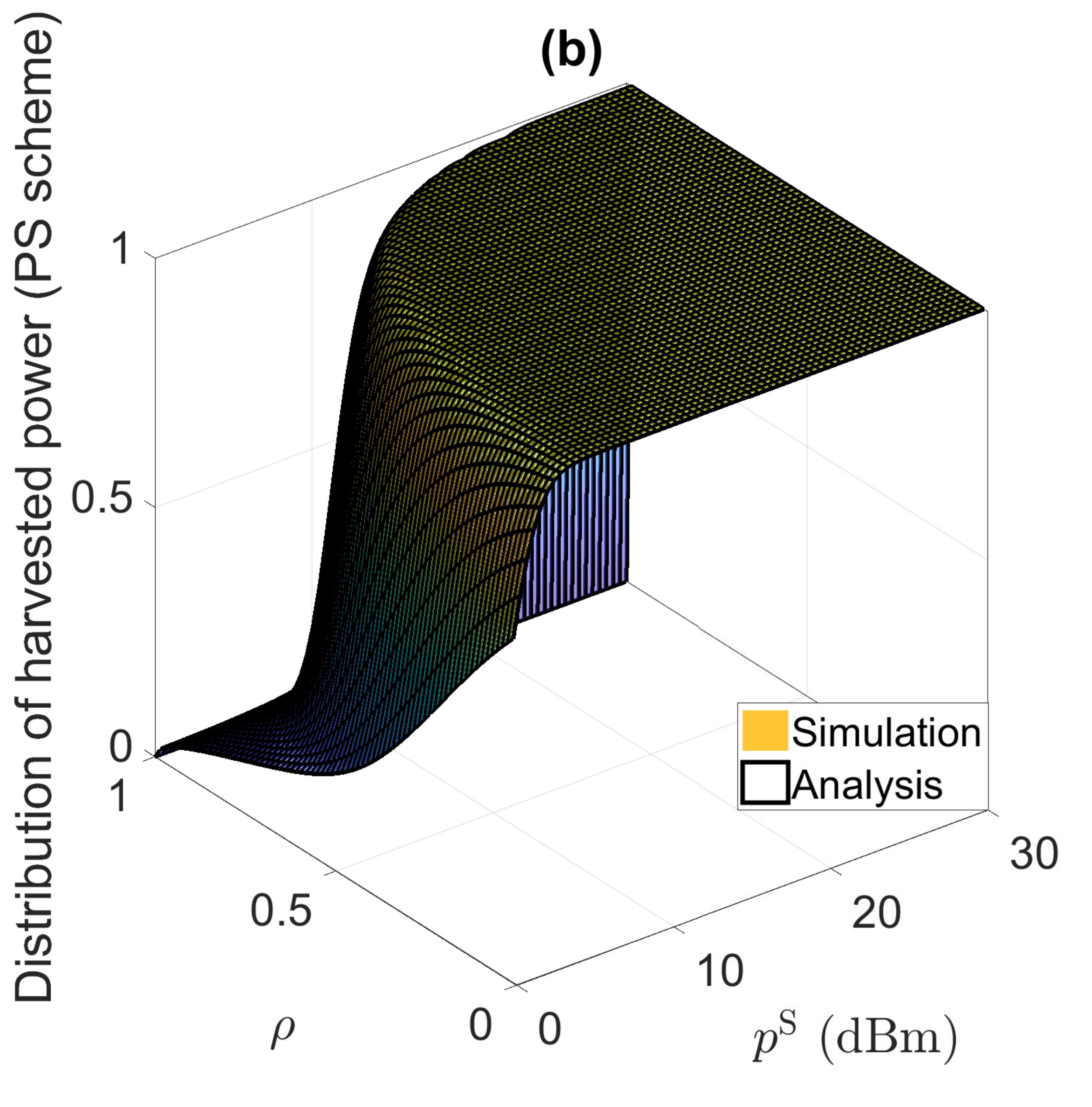}
     \end{minipage}
        \caption{Distribution of (a) SINR and (b) harvested power in PS-based SWIPT scheme.}
        \label{fig:Prob_PS}
\end{figure}

As shown in Fig.~\ref{fig:system_model}(b), the instantaneous received signal power of receiver $d^{\rm{Rx}}$ at the RF energy harvesting circuit will be divided into two parts (i.e., $\left(1 - \rho\right)$ and $\rho$ with $\rho \in \left[0,1\right]$ for energy harvesting and information decoding, respectively). In this case, the instantaneous SINR and instantaneous harvested power at receiver $d^{\rm{Rx}}$ by conditioning on the active status of transmitter $d^{\rm{Tx}}$ (i.e., ${\cal{H}}_1$) are 
\begin{equation}
\left\{\begin{aligned}
{\rm{SINR}}^{\rm{PS}}_{d^{\rm{Rx}}} = & \frac{\rho p^{\rm{S}} g_{d^{\rm{Tx}}d^{\rm{Rx}}} R^{-\alpha}}{\rho \left(I^{\rm{S}}_{d^{\rm{Rx}}} + I^{\rm{C}}_{d^{\rm{Rx}}} + N^{\rm{RF}}\right) + N_{d^{\rm{Rx}}}},\\
{\rm{PH}}^{\rm{PS}}_{d^{\rm{Rx}}} = & \left(1 - \rho\right) \left[p^{\rm{S}} g_{d^{\rm{Tx}}d^{\rm{Rx}}} R^{-\alpha} + I^{\rm{S}}_{d^{\rm{Rx}}} + I^{\rm{C}}_{d^{\rm{Rx}}} + N^{\rm{RF}}\right] + N_{d^{\rm{Rx}}},
\end{aligned}\right.
\end{equation}
respectively. Transmitter $d^{\rm{Tx}}$ transmits a packet of length $l^{\rm{P}}$ within each time slot if it is active. In this case, the transmission rate should be no less than $\frac{l^{\rm{P}}}{\Delta t}$. To ensure that the information can be successfully decoded, the instantaneous SINR at receiver $d^{\rm{Rx}}$ should be no less than $2^{\frac{l^{\rm{P}}}{\Delta t}} - 1$ (i.e., ${\rm{SINR}}^{\rm{PS}}_{d^{\rm{Rx}}} \ge 2^{\frac{l^{\rm{P}}}{\Delta t}} - 1$). Moreover, to maintain the operation of the receiver, receiver $d^{\rm{Rx}}$ needs to harvest power that is no less than ${\rm{PH}}_0$ (i.e., ${\rm{PH}}^{\rm{PS}}_{d^{\rm{Rx}}} \ge {\rm{PH}}_0$). Consequently, we have the distributions of SINR and harvested power as follows:
\begin{equation}\label{eq:Prob_SINR_PS_def}
P^{{\rm{SINR}},{\rm{PS}}}\left(p^{\rm{S}}, \rho\right) = \lim_{T \rightarrow +\infty} \frac{1}{T} \sum_{t = 1}^T {\mathbbm{1}}_{\left\{\left.{\rm{SINR}}^{\rm{PS}}_{d^{\rm{Rx}}} \ge 2^{\frac{l^{\rm{P}}}{\Delta t}} - 1\right|{\cal{H}}_1, t\right\}}= {\mathbb{P}} \left[\left.{\rm{SINR}}^{\rm{PS}}_{d^{\rm{Rx}}} \ge 2^{\frac{l^{\rm{P}}}{\Delta t}} - 1 \right|{\cal{H}}_1\right]
\end{equation}
and 
\begin{equation}\label{eq:Prob_PH_PS_def}
P^{{\rm{PH}},{\rm{PS}}}\left(p^{\rm{S}}, \rho\right) = \lim_{T \rightarrow +\infty} \frac{1}{T} \sum_{t = 1}^T {\mathbbm{1}}_{\left\{\left. {\rm{PH}}^{\rm{PS}}_{d^{\rm{Rx}}} \ge {\rm{PH}}_0 \right|{\cal{H}}_1, t\right\}} = {\mathbb{P}} \left[\left. {\rm{PH}}^{\rm{PS}}_{d^{\rm{Rx}}} \ge {\rm{PH}}_0 \right|{\cal{H}}_1\right],
\end{equation}
as the performance metrics measuring the link reliability and sustainability, respectively, where ${\mathbbm{1}}_{\left\{\left.{\rm{SINR}}^{\rm{PS}}_{d^{\rm{Rx}}} \ge 2^{\frac{l^{\rm{P}}}{\Delta t}} - 1 \right|{\cal{H}}_1, t\right\}}$ is equal to $1$ if ${\rm{SINR}}^{\rm{PS}}_{d^{\rm{Rx}}} \ge 2^{\frac{l^{\rm{P}}}{\Delta t}} - 1 $ conditioning on ${\cal{H}}_1 $ at slot $t$ and $0$ otherwise and ${\mathbbm{1}}_{\left\{\left. {\rm{PH}}^{\rm{PS}}_{d^{\rm{Rx}}} \ge {\rm{PH}}_0 \right|{\cal{H}}_1, t\right\}}$ can be defined similarly. The derivations of $P^{{\rm{SINR}},{\rm{PS}}}\left(p^{\rm{S}}, \rho\right)$ and $P^{{\rm{PH}},{\rm{PS}}}\left(p^{\rm{S}}, \rho\right)$ are shown in Appendices~\ref{app:PS_SINR} and~\ref{app:PS_energy}, respectively, and numerically verified by checking the consistency between the simulation and analytical results in Figs.~\ref{fig:Prob_PS}(a) and~(b), respectively. Note here that the simulation results are obtained by using the Monte Carlo method.

\subsubsection{TS-Based SWIPT Scheme}

\begin{figure}
     \centering
     \begin{minipage}{8cm}
		\centering
		\includegraphics[width=1\textwidth,trim=10 25 20 5,clip]{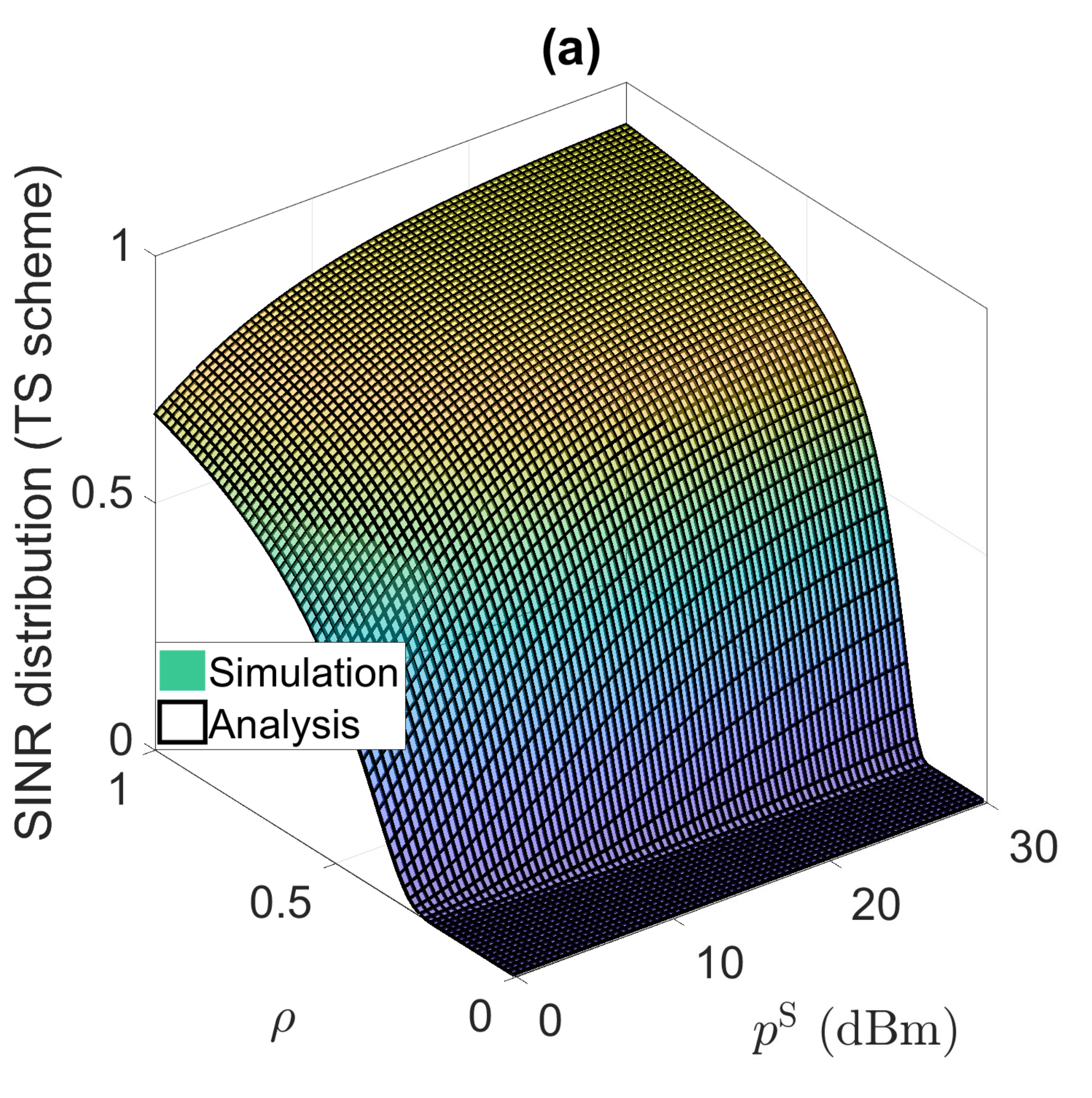}
     \end{minipage}
     \begin{minipage}{8cm}
		\centering
		\includegraphics[width=1\textwidth,trim=10 25 20 5,clip]{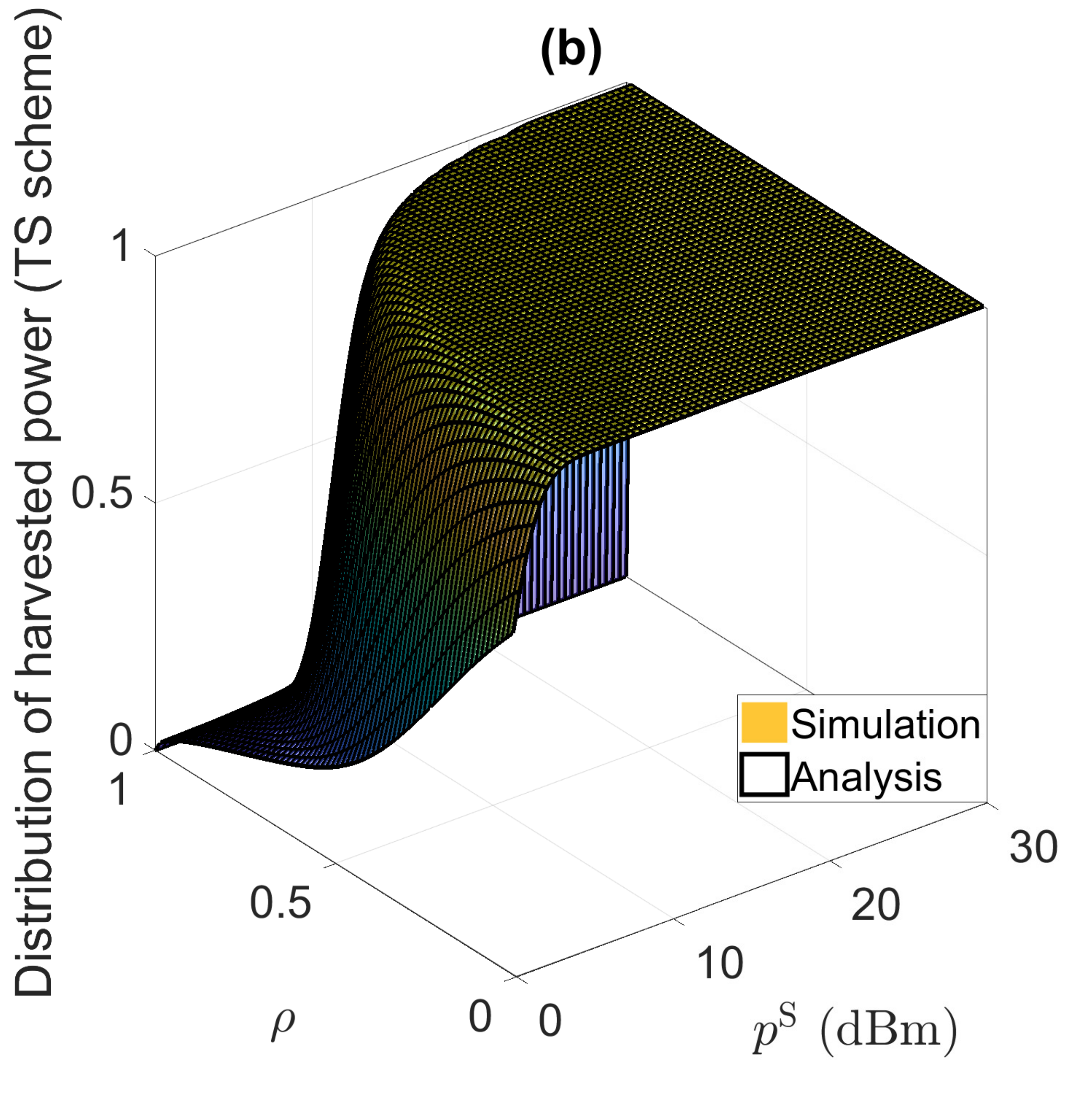}
     \end{minipage}
        \caption{Distribution of (a) SINR and (b) harvested power in TS-based SWIPT scheme.}
        \label{fig:Prob_TS}
\end{figure}

As shown in Fig.~\ref{fig:system_model}(c), the received signal power during $\left(1 - \rho\right)$ and $\rho$ of the slot duration (i.e., $\left(1 - \rho\right) \Delta t$ and $\rho \Delta t$, respectively), will be used for energy harvesting and information decoding, respectively. In this case, the instantaneous SINR and instantaneous harvested power of receiver $d^{\rm{Rx}}$ by conditioning on the active status of transmitter $d^{\rm{Tx}}$ (i.e., ${\cal{H}}_1$) are 
\begin{equation}
\left\{\begin{aligned}
{\rm{SINR}}^{\rm{TS}}_{d^{\rm{Rx}}} = & \frac{p^{\rm{S}} g_{d^{\rm{Tx}}d^{\rm{Rx}}} R^{-\alpha}}{I^{\rm{S}}_{d^{\rm{Rx}}} + I^{\rm{C}}_{d^{\rm{Rx}}} + N^{\rm{RF}} + N_{d^{\rm{Rx}}}},\\
{\rm{\textsc{PH}}}^{\rm{TS}}_{d^{\rm{Rx}}} = & p^{\rm{S}} g_{d^{\rm{Tx}}d^{\rm{Rx}}} R^{-\alpha} + I^{\rm{S}}_{d^{\rm{Rx}}} + I^{\rm{C}}_{d^{\rm{Rx}}} + N^{\rm{RF}} + N_{d^{\rm{Rx}}},
\end{aligned}\right.
\end{equation}
respectively. Similarly, to ensure the successful transmission of a packet, we need to have ${\rm{SINR}}^{\rm{TS}}_{d^{\rm{Rx}}} \ge 2^{\frac{l^{\rm{P}}}{\rho \Delta t}} - 1$ so as to ensure that the cumulative transmitted data within $\rho \Delta t$ is no less than $l^{\rm{P}}$. Moreover, for ease of comparison, we consider that receiver $d^{\rm{Rx}}$ needs to harvest the same amount of energy in the TS-based SWIPT scheme as that in the PS-based SWIPT scheme (i.e., ${\rm{PH}}_0 \Delta t$). In this case, receiver $d^{\rm{Rx}}$ should harvest power that is no less than $\frac{{\rm{PH}}_0 \Delta t}{\left(1 - \rho\right) \Delta t} = \frac{{\rm{PH}}_0}{1 - \rho}$. Consequently, we have the distributions of SINR and harvested power as follows:
\begin{equation}\label{eq:Prob_SINR_TS_def}
P^{{\rm{SINR}},{\rm{TS}}}\left(p^{\rm{S}}, \rho\right) = \lim_{T \rightarrow +\infty} \frac{1}{T} \sum_{t = 1}^T {\mathbbm{1}}_{\left\{\left.{\rm{SINR}}^{\rm{TS}}_{d^{\rm{Rx}}} \ge 2^{\frac{l^{\rm{P}}}{\rho \Delta t}} - 1 \right|{\cal{H}}_1, t\right\}} = {\mathbb{P}} \left[\left.{\rm{SINR}}^{\rm{TS}}_{d^{\rm{Rx}}} \ge 2^{\frac{l^{\rm{P}}}{\rho \Delta t}} - 1 \right|{\cal{H}}_1\right]
\end{equation}
and
\begin{equation}\label{eq:Prob_PH_TS_def}
P^{{\rm{PH}},{\rm{TS}}}\left(p^{\rm{S}}, \rho\right) = \lim_{T \rightarrow +\infty} \frac{1}{T} \sum_{t = 1}^T {\mathbbm{1}}_{\left\{\left. {\rm{PH}}^{\rm{TS}}_{d^{\rm{Rx}}} \ge \frac{{\rm{PH}}_0}{1 - \rho}\right|{\cal{H}}_1, t\right\}} = {\mathbb{P}} \left[\left. {\rm{PH}}^{\rm{TS}}_{d^{\rm{Rx}}} \ge \frac{{\rm{PH}}_0}{1 - \rho}\right|{\cal{H}}_1\right],
\end{equation}
as the performance metrics measuring the link reliability and sustainability, respectively, where ${\mathbbm{1}}_{\left\{\left.{\rm{SINR}}^{\rm{TS}}_{d^{\rm{Rx}}} \ge 2^{\frac{l^{\rm{P}}}{\rho \Delta t}} - 1 \right|{\cal{H}}_1, t\right\}}$ is equal to $1$ if ${\rm{SINR}}^{\rm{TS}}_{d^{\rm{Rx}}} \ge 2^{\frac{l^{\rm{P}}}{\rho \Delta t}} - 1$ conditioning on ${\cal{H}}_1$ at slot $t$ and $0$ otherwise and ${\mathbbm{1}}_{\left\{\left. {\rm{PH}}^{\rm{TS}}_{d^{\rm{Rx}}} \ge \frac{{\rm{PH}}_0}{1 - \rho}\right|{\cal{H}}_1, t\right\}}$ can be defined similarly. The derivations of $P^{{\rm{SINR}},{\rm{TS}}}\left(p^{\rm{S}}, \rho\right)$ and $P^{{\rm{PH}},{\rm{TS}}}\left(p^{\rm{S}}, \rho\right)$ are shown in Appendices~\ref{app:TS_SINR} and~\ref{app:TS_energy}, respectively, and numerically verified in Figs.~\ref{fig:Prob_TS}(a) and~(b), respectively.

\subsection{Transmission Detection Model}

\begin{figure}
     \centering
     \begin{minipage}{8cm}
		\centering
		\includegraphics[width=1\textwidth,trim=10 25 20 5,clip]{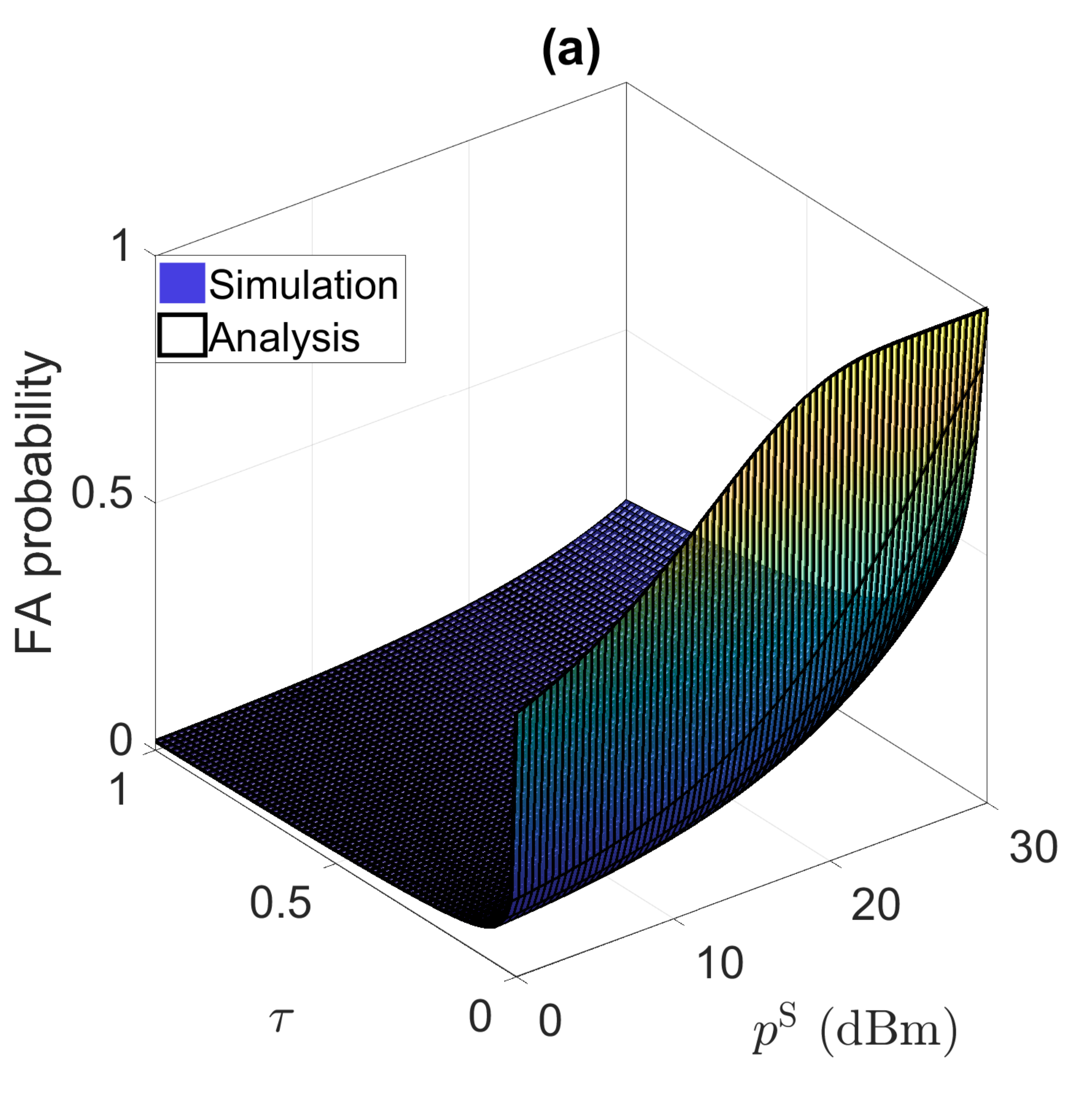}
     \end{minipage}
     \begin{minipage}{8cm}
		\centering
		\includegraphics[width=1\textwidth,trim=10 25 20 5,clip]{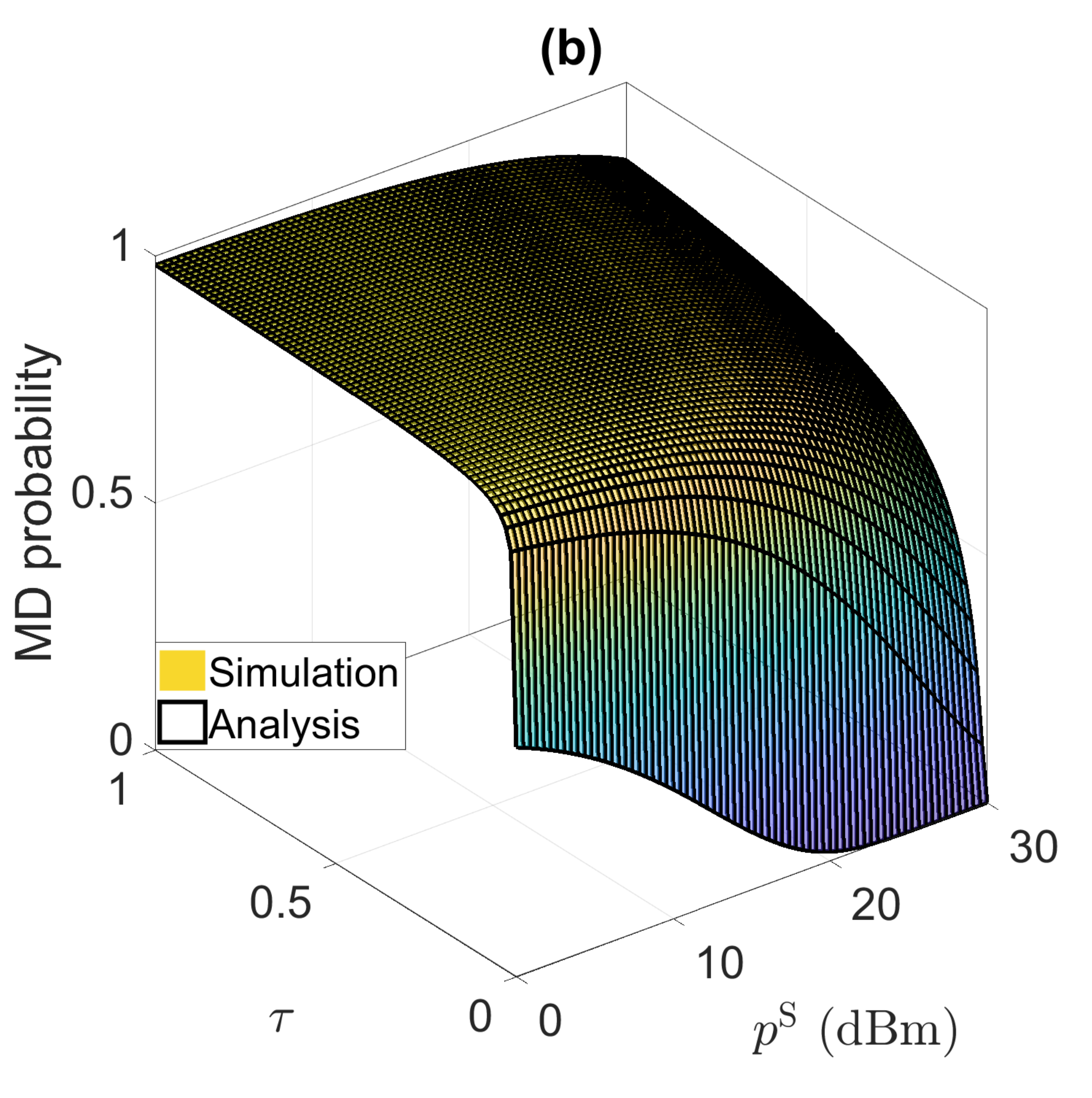}
     \end{minipage}
        \caption{(a) FA probability and (b) MD probability.}
        \label{fig:Prob_FA_MD}
\end{figure}

For the received signal power of adversary $a$, we have the following two cases  based on the activation status of its targeting transmitter (i.e., transmitter $d^{\rm{Tx}}$): 
\begin{enumerate}

\item If transmitter $d^{\rm{Tx}}$ is inactive (i.e., ${\cal{H}}_0$), the instantaneous received signal power of adversary $a$ includes only the interference and noise: 
\begin{equation}
y_a = I^{\rm{S}}_a + I^{\rm{C}}_a + N_a,
\end{equation}
where $I^{\rm{S}}_a = p^{\rm{S}} \sum\limits_{{d^{\rm{Tx}}}' \in \left\{\left.{\cal{D}}^{\rm{Tx}}\backslash \left\{d^{\rm{Tx}}\right\}\right| a \right\}} {\mathbbm{1}}_{{d^{\rm{Tx}}}'} g_{{d^{\rm{Tx}}}'a} \ell\left({\bf{x}}_{{d^{\rm{Tx}}}'}, {\bf{x}}_a\right)$ and $I^{\rm{C}}_a = p^{\rm{C}} \sum\limits_{b \in \left\{\left.{\cal{B}}\right| a \right\}} {\mathbbm{1}}_b h_{b a} \ell\left({\bf{x}}_b, {\bf{x}}_a\right)$ are the interference from other transmitters and that from BSs, respectively, and $N_a$ is the additive noise. Therein, $g_{{d^{\rm{Tx}}}'a} = \left\|{\bf{h}}_{{d^{\rm{Tx}}}'a}^{\rm{H}} {\bf{v}}_{{d^{\rm{Tx}}}'}\right\|^2$ is exponentially distributed~\cite{875282}, $\left\{{\bf{x}}_{{d^{\rm{Tx}}}'}\right\}_{{d^{\rm{Tx}}}' \in \left\{\left.{\cal{D}}^{\rm{Tx}}\backslash \left\{d^{\rm{Tx}}\right\}\right| a \right\}}$ is one of the realizations of the PP of transmitters that excludes transmitter $d^{\rm{Tx}}$ with adversary $a$ as the observation point (i.e., $\Phi_{\left\{\left.{\cal{D}}^{\rm{Tx}}\backslash \left\{d^{\rm{Tx}}\right\}\right| a \right\}}$) and $\left\{{\bf{x}}_b\right\}_{b \in \left\{\left.{\cal{B}}\right| a \right\}}$ is one of the realizations of the PP of BSs with adversary $a$ as the observation point (i.e., $\Phi_{\left\{\left.{\cal{B}}\right| a \right\}}$). 

\item If transmitter $d^{\rm{Tx}}$ is active (i.e., ${\cal{H}}_1$), the instantaneous received signal power of adversary $a$ will additionally include the signal power from transmitter $d^{\rm{Tx}}$ (i.e., $p g_{d^{\rm{Tx}} a} \ell\left({\bf{x}}_{d^{\rm{Tx}}}, {\bf{x}}_a\right)$) and is
\begin{equation}
y_a = p^{\rm{S}} g_{d^{\rm{Tx}} a} \ell\left({\bf{x}}_{d^{\rm{Tx}}}, {\bf{x}}_a\right) + I^{\rm{S}}_a + I^{\rm{C}}_a + N_a,
\end{equation}
where $g_{d^{\rm{Tx}} a} = \left\|{\bf{h}}_{d^{\rm{Tx}} a}^{\rm{H}} {\bf{v}}_{{d^{\rm{Tx}}}}\right\|^2$ is also exponentially distributed~\cite{875282}.
\end{enumerate}
In this case, the adversary can employ binary hypothesis testing to detect the D2D transmission, namely adversary $a$ advocates ${\cal{H}}_0$ (${\cal{H}}_1$) and rejects ${\cal{H}}_1$ (${\cal{H}}_0$) when its instantaneous received signal power (i.e., $y_{a}$) is smaller (larger) than a predetermined threshold $\tau$ (i.e., $y_{a} \mathop{\lessgtr}\limits_{{\cal{H}}_{1}}^{{\cal{H}}_{0}} \tau$). In this framework, adversary $a$ makes a false alarm (FA) when it advocates ${\cal{H}}_1$ (i.e., $y_{a} > \tau$) while ${\cal{H}}_0$ is true, and a miss detection (MD) when it advocates ${\cal{H}}_0$ (i.e., $y_{a} < \tau$) while ${\cal{H}}_1$ is true. After a sufficiently long period (i.e., $T \rightarrow +\infty$), the probabilities with which adversary $a$ makes FA and MD can be defined as follows:
\begin{equation}\label{eq:FA_a}
P^{\rm{FA}}_a\left(p^{\rm{S}}, \tau\right) = \lim_{T \rightarrow +\infty} \frac{1}{T}\sum_{t=1}^T {\mathbbm{1}}_{\left\{\left.y_{a} > \tau \right| {\cal{H}}_{0}, t\right\}} = {\mathbb{P}}\left[\left.y_{a} > \tau \right| {\cal{H}}_{0}\right],
\end{equation}
and 
\begin{equation}\label{eq:MD_a}
P^{\rm{MD}}_a\left(p^{\rm{S}}, \tau\right) = \lim_{T \rightarrow +\infty} \frac{1}{T}\sum_{t=1}^T {\mathbbm{1}}_{\left\{\left.y_{a} < \tau \right|{\cal{H}}_{1}, t\right\}} = {\mathbb{P}}\left[\left.y_{a} < \tau \right| {\cal{H}}_{1}\right],
\end{equation}
respectively, where ${\mathbbm{1}}_{\left\{\left.y_{a} > \tau \right| {\cal{H}}_{0}, t\right\}}$ (${\mathbbm{1}}_{\left\{\left.y_{a} < \tau \right|{\cal{H}}_{1}, t\right\}}$) is equal to $1$ if an FA (MD) occurs at slot $t$ and $0$ otherwise. The derivations of $P^{\rm{FA}}_a\left(p^{\rm{S}}, \tau\right)$ and $P^{\rm{MD}}_a\left(p^{\rm{S}}, \tau\right)$ are shown in Appendices~\ref{app:FA_prob} and~\ref{app:MD_prob}, respectively, and numerically verified in Figs.~\ref{fig:Prob_FA_MD}(a) and~(b), respectively.

%\clearpage

\section{Problem Formulation and Algorithm Design}
\label{sec:game_equilibrium}

Based on the system description in Section~\ref{sec:system_model}, by using game theory, we model the combat between the adversaries and D2D network in Section~\ref{subsec:game_formulation} and then develop an algorithm to obtain their optimal strategies in Sections~\ref{subsec:solution_lower_stage} and~\ref{subsec:solution_upper_stage}.

\subsection{Game Formulation}
\label{subsec:game_formulation}

The interaction (i.e., combat) between the adversaries and D2D network can be modeled in the framework of a two-stage Stackelberg game. The D2D network makes a decision on the SWIPT transmission power $p^{\rm{S}}$ and splitting coefficient $\rho$ first, and the adversaries detect the D2D transmission according to their received signal powers afterward. In this context, the D2D network is the leader at the upper stage while the adversaries are the followers at the lower stage. The Stackelberg game over the representative network nodes (i.e., adversary $a$, transmitter $d^{\rm{Tx}}$, and receiver $d^{\rm{Rx}}$) is a single-leader-single-follower game and defined as follows:
\begin{enumerate}
\item At the lower stage, given the strategy of the D2D network, adversary $a$ aims to minimize its detection error as follows:
\begin{equation}\label{eq:detection_error}
\tau^\star = \arg\min_{\tau} \left\{P^{\rm{FA}}_a \left(p^{\rm{S}}, \tau\right) + P^{\rm{MD}}_a \left(p^{\rm{S}}, \tau\right)\right\}, 
\end{equation}
which is the sum of the probabilities of FA and MD defined in~(\ref{eq:FA_a}) and~(\ref{eq:MD_a}), respectively. It is notable that we omit ${\mathbb{P}}^{{\cal{H}}_1}$ and ${\mathbb{P}}^{{\cal{H}}_0}$ in~(\ref{eq:detection_error}) is due to the following reasons: 1) this is to treat FA and MD equivalently~\cite{6692447}, 2) adversary $a$ is unaware of the active status of transmitter $d^{\rm{Tx}}$ and hence has no information about ${\mathbb{P}}^{{\cal{H}}_1}$ and ${\mathbb{P}}^{{\cal{H}}_0}$, and 3) the detection error defined in~(\ref{eq:detection_error}) is an upper bound of the detection error probability (i.e., $\max\left\{{\mathbb{P}}^{{\cal{H}}_1}, {\mathbb{P}}^{{\cal{H}}_0}\right\}\left[P^{\rm{FA}}_a \left(p^{\rm{S}}, \tau\right) + P^{\rm{MD}}_a \left(p^{\rm{S}}, \tau\right)\right] \ge {\mathbb{P}}^{{\cal{H}}_0} P^{\rm{FA}}_a \left(p^{\rm{S}}, \tau\right) + {\mathbb{P}}^{{\cal{H}}_1} P^{\rm{MD}}_a \left(p^{\rm{S}}, \tau\right)$). 

\item At upper stage, for ${\rm{scheme}} \in \left\{{\rm{TS}}, {\rm{PS}}\right\}$, the network utility maximization problem of transmitter $d^{\rm{Tx}}$ w.r.t. its defending adversary (i.e., adversary $a$) is as follows:
\begin{subequations}\label{eq:upper_problem}
\begin{align}
\max_{p^{\rm{S}}, \rho} & \;\, \left\{u^{\rm{C}} P^{{\rm{SINR}}, {\rm{scheme}}}\left(p^{\rm{S}}, \rho\right) - u^{\rm{P}} \rho p^{\rm{S}}\right\} \label{eq:upper_problem_obj}\\
{\text{s.\,t.}} & \;\, P^{{\rm{PH}}, {\rm{scheme}}}\left(p^{\rm{S}}, \rho\right) \ge 1 - \varepsilon^{\rm{P}} \label{eq:upper_problem_constr_PH}\\
& \, P^{\rm{FA}}_{a} \left(p^{\rm{S}},\tau^\star\right) + P^{\rm{MD}}_{a} \left(p^{\rm{S}},\tau^\star\right) \ge 1 - \varepsilon^{\rm{C}} \label{eq:upper_problem_constr_CC}\\
& \, p^{\rm{S}} \in \left[{\underline{p^{\rm{S}}}}, {\overline{p^{\rm{S}}}}\right], \, \rho \in \left[\epsilon, 1\right].
\end{align}
\end{subequations}
Therein, the objective (i.e.,~(\ref{eq:upper_problem_obj})) is the network utility defined as the difference between the reward (i.e., $u^{\rm{C}}$) obtained by ensuring the link reliability (i.e., $P^{{\rm{SINR}},{\rm{scheme}}}\left(p^{\rm{S}}, \rho\right)$ for ${\rm{scheme}} \in \left\{{\rm{TS}}, {\rm{PS}}\right\}$) and the power cost incurred by the wireless communication (i.e., $u^{\rm{P}} \rho p^{\rm{S}}$ with $u^{\rm{P}}$ as the unit cost). The power consumption due to the wireless transmission is added to that incurred by the wireless power transfer for supporting the system operation of the receivers. The first constraint, i.e.,~(\ref{eq:upper_problem_constr_PH}), is to ensure that the power outage probability at receiver $d^{\rm{Rx}}$ is no more than a certain threshold (i.e., $\varepsilon^{\rm{P}}$). The second constraint, i.e.,~(\ref{eq:upper_problem_constr_CC}), is to ensure a certain level (i.e., $1 - \varepsilon^{\rm{C}}$) for the communication covertness, where $\tau^\star$ is the best response from the lower stage (i.e.,~(\ref{eq:detection_error})). ${\underline{p^{\rm{S}}}}$ ($\epsilon$) and ${\overline{p^{\rm{S}}}}$ ($1$) are the lower and upper bounds of $p^{\rm{S}}$ ($\rho$), respectively.
\end{enumerate}

\subsection{Solution to Lower-Stage Detection Minimization Problem}
\label{subsec:solution_lower_stage}

Given SWIPT transmission power $p^{\rm{S}}$, we analyze the characteristics of the lower-stage detection minimization problem for adversary $a$. The detection error of adversary $a$ (i.e.,~(\ref{eq:detection_error})) can be rewritten as follows:
\begin{equation}
\begin{aligned}
& P^{\rm{FA}}_a\left(p^{\rm{S}}, \tau\right) + P^{\rm{MD}}_a\left(p^{\rm{S}}, \tau\right) \\
= & 1 - F_{I^{\rm{S}}_a + I^{\rm{C}}_a} \left(\tau - N_a\right) + \int^{\tau - N_a}_0 f_{I^{\rm{S}}_a + I^{\rm{C}}_a}\left(t\right) F_{p^{\rm{S}} g_{d^{\rm{Tx}} a} \ell\left({\bf{x}}_{d^{\rm{Tx}}}, {\bf{x}}_a\right)} \left(\tau - N_a - t\right) {\rm{d}} t\\
= & 1 - \int^{\tau - N_a}_0 f_{I^{\rm{S}}_a + I^{\rm{C}}_a}\left(t\right) \left[1 - F_{p^{\rm{S}} g_{d^{\rm{Tx}} a} \ell\left({\bf{x}}_{d^{\rm{Tx}}}, {\bf{x}}_a\right)} \left(\tau - N_a - t\right)\right] {\rm{d}} t,
\end{aligned}
\end{equation}
and its first-order derivative is 
\begin{equation}\label{eq:first_derivative_detection_error}
\begin{aligned}
&\frac{{\rm{d}}}{{\rm{d}} \tau} \left[P^{\rm{FA}}_a\left(p^{\rm{S}}, \tau\right) + P^{\rm{MD}}_a\left(p^{\rm{S}}, \tau\right)\right] \\
= & -f_{I^{\rm{S}}_a + I^{\rm{C}}_a}\left(\tau - N_a\right) + \int^{\tau - N_a}_0 f_{I^{\rm{S}}_a + I^{\rm{C}}_a}\left(t\right) \frac{{\rm{d}}}{{\rm{d}} \tau} F_{p^{\rm{S}} g_{d^{\rm{Tx}} a} \ell\left({\bf{x}}_{d^{\rm{Tx}}}, {\bf{x}}_a\right)} \left(\tau - N_a - t\right) {\rm{d}} t.
\end{aligned}
\end{equation}
There are two notable properties regarding the first-order derivative of the detection error of adversary $a$ (i.e., $\frac{{\rm{d}}}{{\rm{d}} \tau} \left[P^{\rm{FA}}_a\left(p^{\rm{S}}, \tau\right) + P^{\rm{MD}}_a\left(p^{\rm{S}}, \tau\right)\right]$ in~(\ref{eq:first_derivative_detection_error})). First, when $\tau$ gradually approaches the right-hand side of $N_a$ (i.e., $\tau \rightarrow N_a^+$), the first-order derivative of the detection error of adversary $a$ is smaller than $0$ as
\begin{equation}\label{eq:first_derivative_detection_error_left_side}
\lim_{\tau \rightarrow N_a^+} \frac{{\rm{d}}}{{\rm{d}} \tau} \left[P^{\rm{FA}}_a\left(p^{\rm{S}}, \tau\right) + P^{\rm{MD}}_a\left(p^{\rm{S}}, \tau\right)\right] = -f_{I^{\rm{S}}_a + I^{\rm{C}}_a}\left(0^+\right) \le 0,
\end{equation}
where $f_{I^{\rm{S}}_a + I^{\rm{C}}_a}\left(\cdot\right)$ is a probability density function (PDF) and first-order derivation of~(\ref{eq:inverse_laplace_laplace_I_a}). Second, when $\tau$ approaches $+\infty$, the first-order derivative of the detection error of adversary $a$ is larger than $0$ as
\begin{equation}\label{eq:first_derivative_detection_error_right_side}
\begin{aligned}
& \lim_{\tau \rightarrow +\infty} \frac{{\rm{d}}}{{\rm{d}} \tau} \left[P^{\rm{FA}}_a\left(p^{\rm{S}}, \tau\right) + P^{\rm{MD}}_a\left(p^{\rm{S}}, \tau\right)\right] \\
= & \underbrace{\lim_{\tau \rightarrow +\infty} -f_{I^{\rm{S}}_a + I^{\rm{C}}_a}\left(\tau - N_a\right)}_{=0} + \lim_{\tau \rightarrow +\infty} \int^{\tau - N_a}_0 \underbrace{f_{I^{\rm{S}}_a + I^{\rm{C}}_a}\left(t\right)}_{\ge 0} \underbrace{\frac{{\rm{d}}}{{\rm{d}} \tau} F_{p^{\rm{S}} g_{d^{\rm{Tx}} a} \ell\left({\bf{x}}_{d^{\rm{Tx}}}, {\bf{x}}_a\right)} \left(\tau - N_a - t\right)}_{\mathop \ge \limits^{(a)} 0} {\rm{d}} t \ge 0,
\end{aligned}
\end{equation}
where (a) follows
\begin{equation}\label{eq:PDF_pg_dTxa_l_substituted}
\frac{{\rm{d}}}{{\rm{d}} \tau} F_{p^{\rm{S}} g_{d^{\rm{Tx}} a} \ell\left({\bf{x}}_{d^{\rm{Tx}}}, {\bf{x}}_a\right)} \left(\tau - N_a - t\right) \mathop = \limits^{(\ref{eq:PDF_pg_dTxa_l})} \int^\infty_0 f_{r_{d^{\rm{Tx}} a}}\left(r\right) \exp\left(-\frac{\left(\tau - N_a - t\right) r^{\alpha}}{p^{\rm{S}}}\right) \frac{r^{\alpha}}{p^{\rm{S}}} {\rm{d}} r \ge 0.
\end{equation}
These two notable properties of $\frac{{\rm{d}}}{{\rm{d}} \tau} \left[P^{\rm{FA}}_a\left(p^{\rm{S}}, \tau\right) + P^{\rm{MD}}_a\left(p^{\rm{S}}, \tau\right)\right]$ in~(\ref{eq:first_derivative_detection_error}) (i.e.,~(\ref{eq:first_derivative_detection_error_left_side}) and~(\ref{eq:first_derivative_detection_error_right_side})) together with its continuity w.r.t $\tau \in \left[N_a^+, +\infty\right)$ indicates that there exists at least one $\tau^\star \in \left[N_a^+, +\infty\right)$ such that $\left. \frac{{\rm{d}}}{{\rm{d}} \tau} \left[P^{\rm{FA}}_a\left(p^{\rm{S}}, \tau\right) + P^{\rm{MD}}_a\left(p^{\rm{S}}, \tau\right)\right]\right|_{\tau = \tau^\star} = 0$ according to the intermediate value theorem~\cite{cates2019cauchy}. Consequently, there exists at least one $\tau^\star \in \left[N_a^+, +\infty\right)$ that minimizes the detection error for adversary $a$ given any SWIPT transmission power $p^{\rm{S}}$. 

\begin{figure}
		\centering
		\includegraphics[width=0.5\textwidth,trim=10 10 70 60,clip]{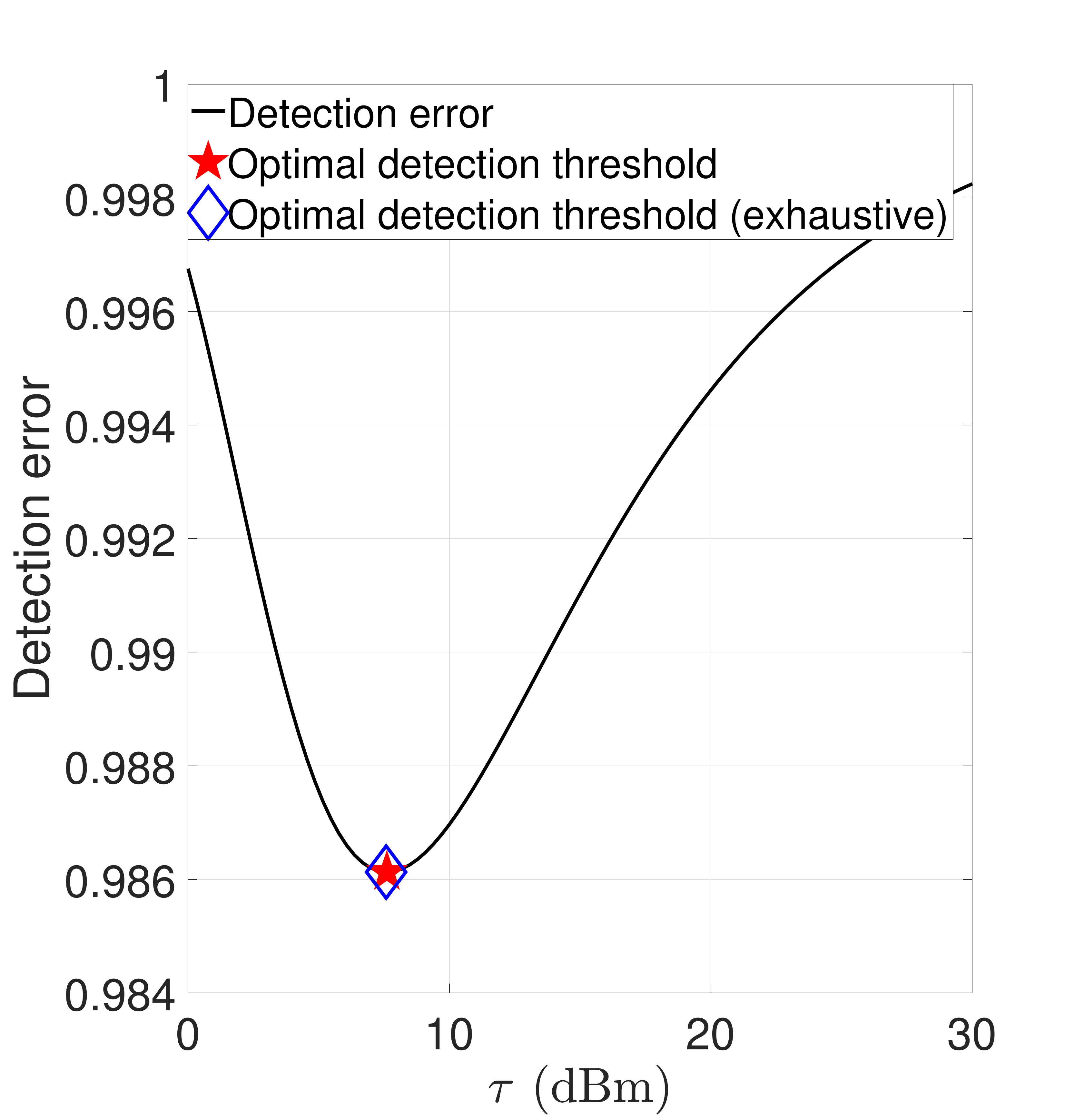}
        \caption{The existence and uniqueness of the solution to the lower-stage problem with $p^{\rm{S}} = 10\, {\rm{dBm}}$.}
        \label{fig:adversary_uniqueness}
\end{figure}

As $F_{p^{\rm{S}} g_{d^{\rm{Tx}} a} \ell\left({\bf{x}}_{d^{\rm{Tx}}}, {\bf{x}}_a\right)} \left(\cdot\right)$ is the cumulative distribution function (CDF) of $p^{\rm{S}} g_{d^{\rm{Tx}} a} \ell\left({\bf{x}}_{d^{\rm{Tx}}}, {\bf{x}}_a\right)$, it will gradually approach and finally stop at its probabilistic upper bound (i.e., $1$). This means that there exists a $\tau_0 > N_a$ such that $F_{p^{\rm{S}} g_{d^{\rm{Tx}} a} \ell\left({\bf{x}}_{d^{\rm{Tx}}}, {\bf{x}}_a\right)} \left(\tau\right) = 1$ when $\tau > \tau_0$. In this case, when $\tau < \tau_0$, 
\begin{equation}
P^{\rm{FA}}_a\left(p^{\rm{S}}, \tau\right) + P^{\rm{MD}}_a\left(p^{\rm{S}}, \tau\right) = 1 - \int^{\tau - N_a}_0 \underbrace{f_{I^{\rm{S}}_a + I^{\rm{C}}_a}\left(\tau - N_a - t\right)}_{>0} \underbrace{\left[1 - F_{p^{\rm{S}} g_{d^{\rm{Tx}} a} \ell\left({\bf{x}}_{d^{\rm{Tx}}}, {\bf{x}}_a\right)} \left(t\right)\right]}_{>0} {\rm{d}} t
\end{equation}
decreases w.r.t. $\tau$. Later, when $\tau > \tau_0$,
\begin{equation}
\begin{aligned}
P^{\rm{FA}}_a\left(p^{\rm{S}}, \tau\right) + P^{\rm{MD}}_a\left(p^{\rm{S}}, \tau\right) = & 1 - \int^{\tau_0 - N_a}_0 \underbrace{f_{I^{\rm{S}}_a + I^{\rm{C}}_a}\left(\tau - N_a - t\right)}_{>0} \underbrace{\left[1 - F_{p^{\rm{S}} g_{d^{\rm{Tx}} a} \ell\left({\bf{x}}_{d^{\rm{Tx}}}, {\bf{x}}_a\right)} \left(t\right)\right]}_{>0} {\rm{d}} t \\
& - \int^{\tau - N_a}_{\tau_0 - N_a} \underbrace{f_{I^{\rm{S}}_a + I^{\rm{C}}_a}\left(\tau - N_a - t\right)}_{>0} \underbrace{\left[1 - F_{p^{\rm{S}} g_{d^{\rm{Tx}} a} \ell\left({\bf{x}}_{d^{\rm{Tx}}}, {\bf{x}}_a\right)} \left(t\right)\right]}_{=0} {\rm{d}} t\\
= & 1 - \int^{\tau_0 - N_a}_0 \underbrace{f_{I^{\rm{S}}_a + I^{\rm{C}}_a}\left(\tau - N_a - t\right)}_{>0} \underbrace{\left[1 - F_{p^{\rm{S}} g_{d^{\rm{Tx}} a} \ell\left({\bf{x}}_{d^{\rm{Tx}}}, {\bf{x}}_a\right)} \left(t\right)\right]}_{>0} {\rm{d}} t.
\end{aligned}
\end{equation}
Here, as $\int^{\tau_0 - N_a}_0 \underbrace{f_{I^{\rm{S}}_a + I^{\rm{C}}_a}\left(\tau - N_a - t\right)}_{>0} \underbrace{\left[1 - F_{p^{\rm{S}} g_{d^{\rm{Tx}} a} \ell\left({\bf{x}}_{d^{\rm{Tx}}}, {\bf{x}}_a\right)} \left(t\right)\right]}_{>0} {\rm{d}} t$ decreases w.r.t $\tau$, $P^{\rm{FA}}_a\left(p^{\rm{S}}, \tau\right) + P^{\rm{MD}}_a\left(p^{\rm{S}}, \tau\right)$ will increase. Consequently, the detection error of adversary $a$ will decrease first and increase later, which induces a unimodal function and admits a global minimum as shown in Fig.~\ref{fig:adversary_uniqueness}. This means that the aforementioned $\tau^\star$ is unique and globally minimizes the detection error for adversary $a$, which is thereby the global optimal detection threshold and regarded as the best response from the lower stage. Regarding an unconstrained optimization problem with the objective as a unimodal function, Rosenbrock method is applicable to find the optimal solution, which is a numerical optimization algorithm that requires to evaluate only the objective function instead of the gradient of the objective function and hence is inexpensive in implementation~\cite{bazaraa2013nonlinear}. The analysis of the convergence and complexity of Rosenbrock method can be found in~\cite{bazaraa2013nonlinear}. The optimality can be verified by the consistency between the optimal solution obtained by Rosenbrock method (i.e., the optimal detection threshold in Fig~\ref{fig:adversary_uniqueness}) and the optimal solution obtained by exhaustive search (i.e., detection threshold (exhaustive) in Fig.~\ref{fig:adversary_uniqueness}). The exhaustive search finds the optimal solution via: 1) discretizing the domain of definition, 2) sequentially evaluating the objective over these discretized points, and 3) outputting the point with the lowest objective function. 

\subsection{Solution to Upper-Stage Network Utility Maximization Problem}
\label{subsec:solution_upper_stage}

The network utility maximization defined in the upper stage (i.e.,~(\ref{eq:upper_problem})) contains multiple integrals in both its objective and constraints. In this case, it is very difficult to verify the convexity of the objective and constraints and further the existence and uniqueness of the solution to the upper-stage problem. Consequently, it is difficult to apply convex optimization methods to solve it. To address this challenge, we use GA~\cite{OLIVETO201521} and take into account the best response from the lower stage to obtain the solution to the upper-stage problem. The GA is an evolutionary algorithm and aims to employ and thereafter evolve a population rather than an individual to find the optimal solution of a nonconvex problem~\cite{OLIVETO201521}. This will significantly enhance the probability of finding the globally optimal solution. The analysis of the optimality, convergence, and complexity of GA can be found in~\cite{OLIVETO201521}.

\section{Performance Evaluation}
\label{sec:numerical_results}

\begin{table*}[!]
	\centering
	\caption{Symbols and descriptions}
	\begin{tabular}{|c|l|c|}
		\hline
		\hline
		{\bf{Symbol}} & {\bf{Description}} & {\bf{Default value}}\\
		\hline
		\makecell[c]{$d^{\rm{Tx}}$, $d^{\rm{Rx}}$, $a$,\\ $b$, $u$} & \makecell[l]{Indexes of the typical transmitter,  typical receiver, nearest \\adversary of D2D transmitter $d^{\rm{Tx}}$, BS, and CU, respectively.} & --- \\
        \hline
		${\cal{D}}^{\rm{Tx}}$, ${\cal{A}}$, ${\cal{B}}$, ${\cal{U}}$ & Sets of the transmitters, adversaries, BSs, and CUs, respectively. & --- \\
        \hline
		\makecell[c]{$\Phi_{{\cal{D}}^{\rm{Tx}}}$, $\Phi_{\cal{A}}$, \\$\Phi_{\cal{B}}$, $\Phi_{\cal{U}}$} & PPs of transmitters, adversaries, BSs, and CUs, respectively. & --- \\
		\hline
		\makecell[c]{$\lambda_{{\cal{D}}^{\rm{Tx}}}$, $\lambda_{\cal{A}}$, \\$\lambda_{\cal{B}}$, $\lambda_{\cal{U}}$} & Densities of $\Phi_{{\cal{D}}^{\rm{Tx}}}$, $\Phi_{\cal{A}}$, $\Phi_{\cal{B}}$, and $\Phi_{\cal{U}}$, respectively. & \makecell[c]{$0.1\slash {\rm{m}}^2$, $0.002\slash {\rm{m}}^2$, \\$0.01\slash {\rm{m}}^2$, $0.1\slash {\rm{m}}^2$.}\\	
		\hline
		${\cal{H}}_0$, ${\cal{H}}_1$, ${\cal{C}}_0$, ${\cal{C}}_1$ & \makecell[l]{Events that the transmitter is inactive and active, and the BS is \\inactive and active, respectively.} & --- \\	
        \hline
		${\mathbb{P}}^{{\cal{H}}_0}$, ${\mathbb{P}}^{{\cal{H}}_1}$, ${\mathbb{P}}^{{\cal{C}}_0}$, ${\mathbb{P}}^{{\cal{C}}_1}$ & Probabilities of events ${\cal{H}}_0$, ${\cal{H}}_1$, ${\cal{C}}_0$, and ${\cal{C}}_1$, respectively. & $0.5$, $0.5$, $0.5$, $0.5$. \\	
		\hline
		$p^{\rm{S}}$, $p^{\rm{C}}$, $\rho$, $\tau$ & \makecell[l]{SWIPT transmission power, cellular transmission power, splitting\\ coefficient, and power detection threshold, respectively.} & \makecell[c]{$\left[0, 30\right] {\rm{dBm}}$, $30\,{\rm{dBm}}$, \\$\left[0.01, 1\right]$, ---.} \\
		\hline
		\makecell[c]{$g_{d^{\rm{Tx}} d^{\rm{Rx}}}$, $R$, \\$h_{b d^{\rm{Rx}}}$, ${\bf{x}}_{d^{\rm{Tx}}}$} & \makecell[l]{Channel and distance between transmitter $d^{\rm{Tx}}$ and receiver $d^{\rm{Rx}}$, \\small-scale fading between BS $b$ and receiver $d^{\rm{Tx}}$, and location \\of transmitter $d^{\rm{Tx}}$, respectively.} & \makecell[c]{---, $1\, {\rm{m}}$, \\$\exp\left(1\right)$, ---.}\\
		\hline
		$M$, $\alpha$, $\varepsilon^{\rm{C}}$, $\varepsilon^{\rm{P}}$ & \makecell[l]{Number of antennas, path-loss exponent, detection error \\threshold, and power outage probability threshold, respectively.} & \makecell[c]{$10$, $4$, \\$0.01$, $0.01$.} \\
		\hline
		$N_a$, $N^{\rm{RF}}$, $N_{d^{\rm{Rx}}}$ & \makecell[l]{Noises at adversary $a$, RF energy harvesting circuit, and \\receiver $d^{\rm{Rx}}$, respectively.} & $- 90\,{\rm{dBm}}$. \\
		\hline
		$I^{\rm{S}}_{d^{\rm{Rx}}}$, $I^{\rm{S}}_a$, $I^{\rm{C}}_{d^{\rm{Rx}}}$, $I^{\rm{C}}_a$ & \makecell[l]{Sum of interference from transmitters to receiver $d^{\rm{Rx}}$, that to \\adversary $a$, sum of interference from BSs to receiver $d^{\rm{Rx}}$, and \\that to adversary $a$, respectively.} & --- \\
		\hline
		$u^{\rm{C}}$, $u^{\rm{P}}$ & Reward of link reliability and power cost, respectively. & $1$, $1$.\\
		\hline
		$l^{\rm{P}}$, $\Delta t$, ${\rm{PH}}_0$ & \makecell[l]{Packet size, slot duration, and threshold of harvested power, \\respectively.} & \makecell[c]{$2\,{\rm{bits}}\slash {\rm{Hz}}$, $1\,{\rm{second}}$, \\$10\,{\rm{dBm}}$.}\\
		\hline
		$f_{\nu}$, $F_{\nu}$, ${\mathcal{L}}_{\nu}$, ${\mathcal{L}}^{-1}_{\nu}$ & \makecell[l]{PDF, CDF, Laplace transfer, and inverse Laplace transforms \\of ${\nu}$, respectively.} & --- \\
		\hline
		${\mathbb{P}}$, ${\mathbb{E}}$ & Probabilistic and expectation operators, respectively. & --- \\
		\hline
	\end{tabular}
	\label{tab:notation_value}
\end{table*}

This section presents the numerical results to evaluate the system performance of the SWIPT-enabled D2D network. The parameter setting is given in Table~\ref{tab:notation_value} together with their descriptions. First, we investigate the Stackelberg equilibrium in Fig.~\ref{fig:stackelberg_equilibrium}. Second, we evaluate the impact of detection error threshold $\varepsilon^{\rm{C}}$ and power outage probability threshold $\varepsilon^{\rm{P}}$ on the network utility in Fig.~\ref{fig:impact_varepsilon_C_P}. Third, we evaluate the gain obtained from reducing the D2D transmission distance and the advantages of the PS-based SWIPT scheme compared with the TS-based SWIPT scheme regarding the threat from the adversaries in Fig.~\ref{fig:impact_R_lambda_A}. Fourth, we further investigate the role of the cellular network in improving the SWIPT-enabled D2D network in Fig.~\ref{fig:impact_power_C_lambda_A}. Fifth, we study the impact of the threshold of harvested power and cellular transmission power in Fig.~\ref{fig:impact_PH0_power_C}. 

%\clearpage

\subsection{Stackelberg Equilibrium}
\label{subsec:stackelberg_equilibrium}

\begin{figure}
     \centering
     \begin{minipage}{8cm}
		\centering
		\includegraphics[width=1\textwidth,trim=5 5 20 15,clip]{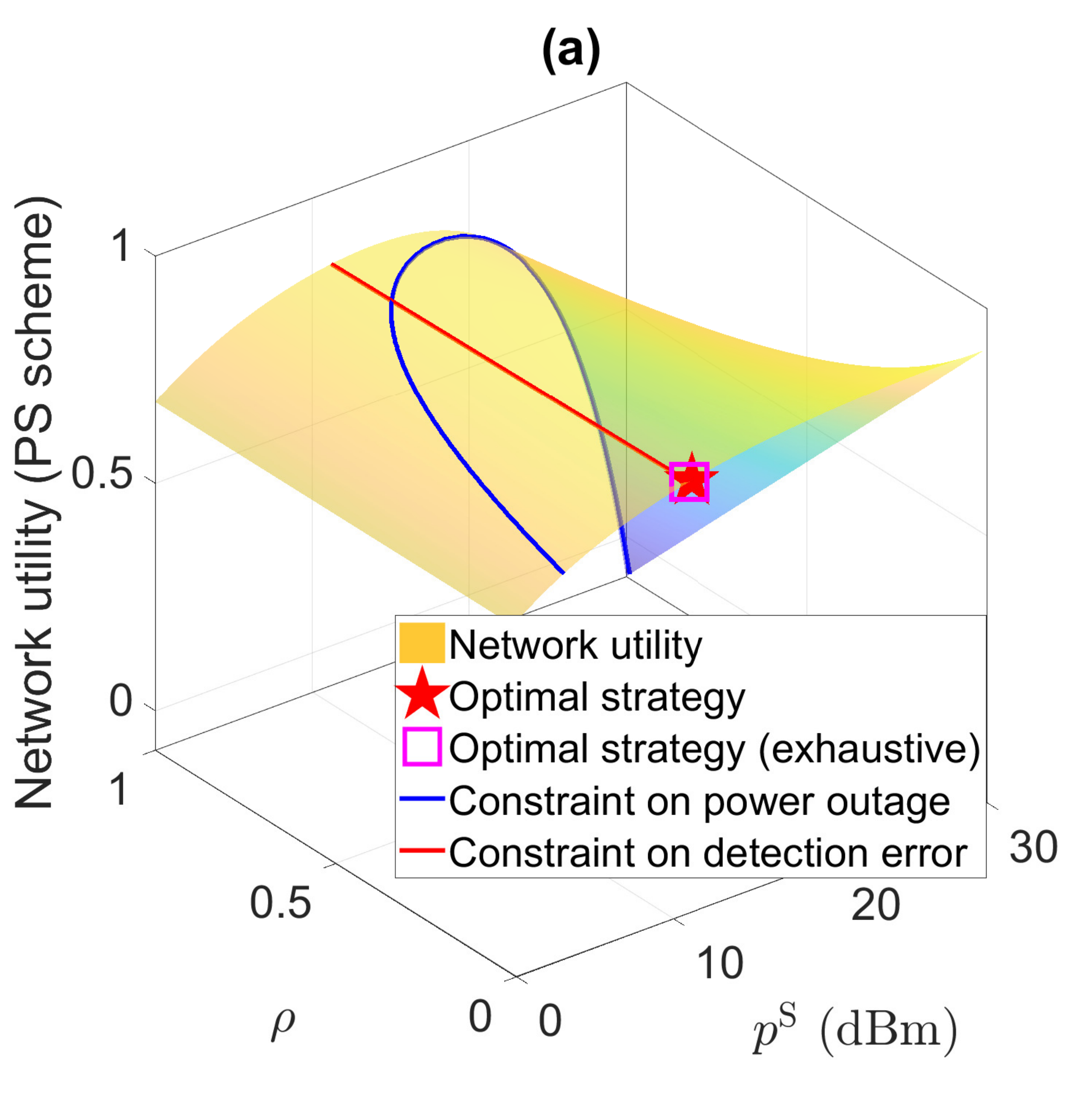}
     \end{minipage}
     \begin{minipage}{8cm}
		\centering
		\includegraphics[width=1\textwidth,trim=5 5 20 15,clip]{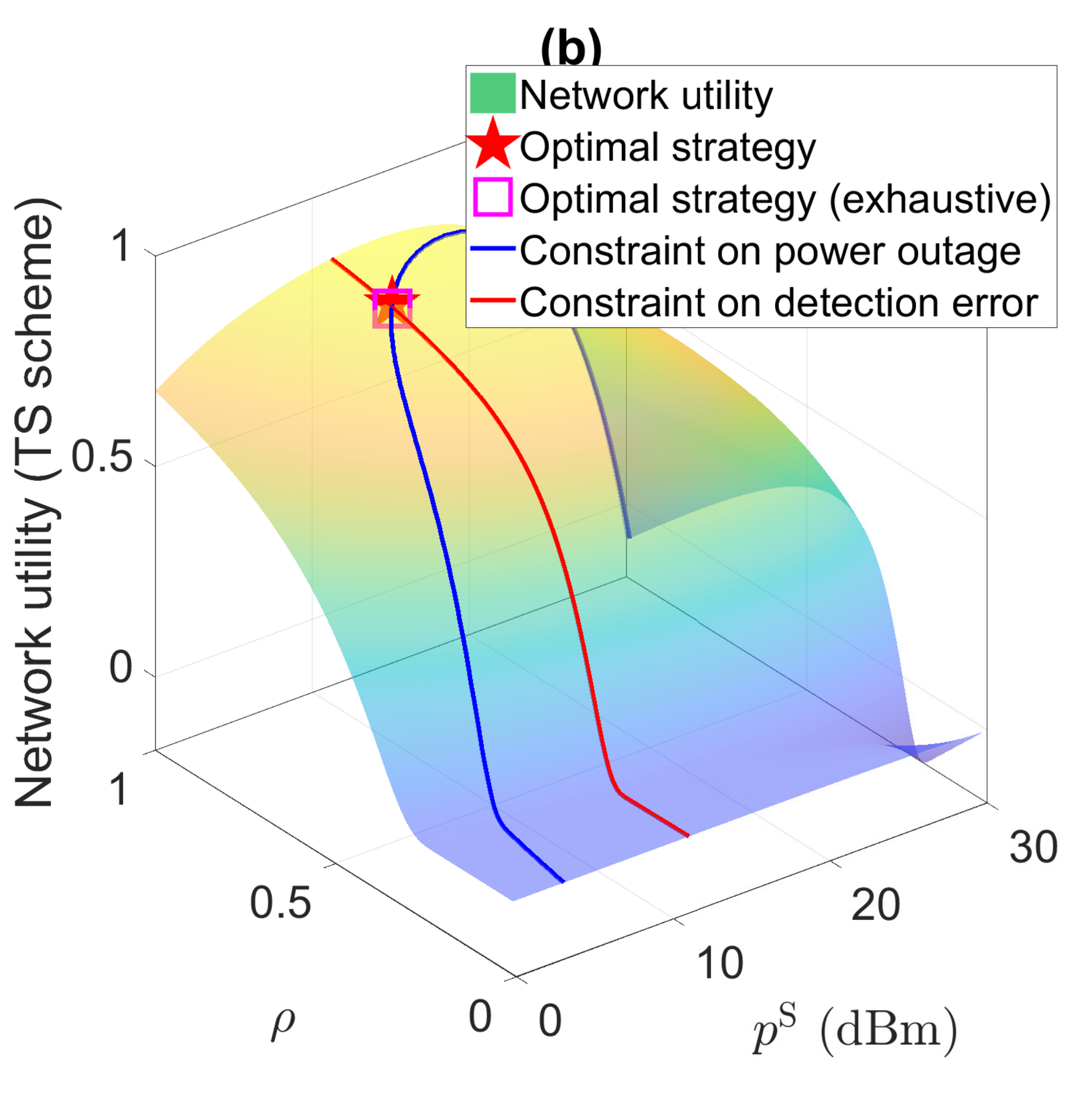}
     \end{minipage}
        \caption{Stackelberg equilibrium in (a) PS and (b) TS-based SWIPT schemes.}
        \label{fig:stackelberg_equilibrium}
\end{figure}

Figure~\ref{fig:stackelberg_equilibrium} depicts the network utility as a function of the splitting coefficient $\rho$ and SWIPT transmission power $p^{\rm{S}}$. In Fig.~\ref{fig:stackelberg_equilibrium}, we can observe that the optimal strategy (i.e., optimal $\rho$ and $p^{\rm{S}}$) maximizes the network utility within the feasible domain, and the boundary of which is determined by the constraints on power outage and detection error (i.e.,~(\ref{eq:upper_problem_constr_PH}) and~(\ref{eq:upper_problem_constr_CC}), respectively). The optimal strategy is obtained by the developed bi-level algorithm and is consistent with the optimal strategy obtained by exhaustive search (i.e., optimal strategy (exhaustive)). In this case, the optimality of the strategy obtained by the bi-level algorithm is validated. Moreover, the network utility is maximized over different strategies in different SWIPT schemes. First, in the PS-based SWIPT scheme, the network utility is maximized when $\rho$ stays at its lower bound and $p^{\rm{S}}$ is around $10$\,dBm. This means that the network utility can be maximized even when only tiny amount of the received signal power is used for information decoding, which implies that most of the received signal power will be used for energy harvesting and later to support the system operation of the receiver. In this case, the power outage probability at the receiver is low and high network sustainability is achieved. On the other hand, in the TS-based SWIPT scheme, the network utility is maximized when $\rho$ is approximately $0.8$ and $p^{\rm{S}}$ is around $10$\,dBm. In this case, most of the received signal power will be used for information decoding, which means that the power outage probability at the receiver is high and further the network sustainability is difficult to be guaranteed. Based on the above discussion, the PS-based SWIPT scheme outperforms the TS-based SWIPT scheme in terms of network sustainability. 

%\clearpage

\subsection{Impact of Constraints on Power Outage and Detection Error}

\begin{figure}
		\centering
		\includegraphics[width=0.5\textwidth,trim=10 5 35 40, clip]{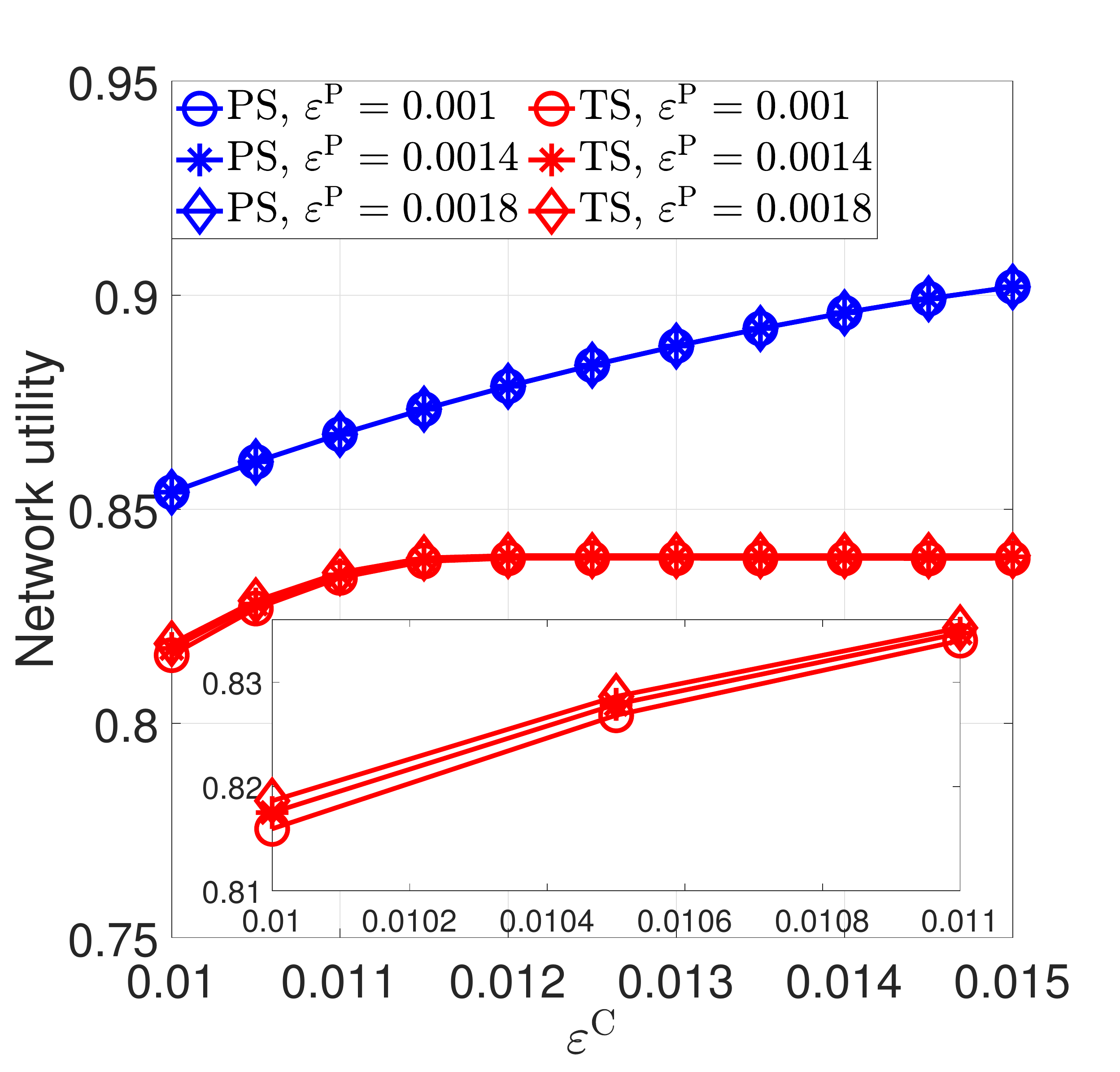}
        \caption{Impact of the constraints on power outage and detection error.}
        \label{fig:impact_varepsilon_C_P}
\end{figure}

In Fig.~\ref{fig:impact_varepsilon_C_P}, we evaluate the impact of the power outage probability threshold $\varepsilon^{\rm{P}}$ and detection error threshold $\varepsilon^{\rm{C}}$ on network utility. It can be observed again that the PS-based SWIPT scheme outperforms the TS-based SWIPT scheme in terms of network utility. This can be explained similarly as that in Section~\ref{subsec:stackelberg_equilibrium} that the PS-based SWIPT scheme requires a small amount of received signal power for information decoding while the TS-based SWIPT scheme will consume a large amount of the received signal power for information decoding. In this case, most of the received signal power in the PS-based SWIPT scheme, while only a small amount of that in the TS-based SWIPT scheme, will be used for energy harvesting. This also well explains why the constraint on power outage (i.e., $\varepsilon^{\rm{P}}$ in~(\ref{eq:upper_problem_constr_PH})) does not affect network utility of the PS-based SWIPT scheme while affecting that of the TS-based SWIPT scheme. Another interesting result is that the constraint on detection error (i.e., $\varepsilon^{\rm{C}}$ in~(\ref{eq:upper_problem_constr_CC})) significantly affects network utility in the PS-based SWIPT scheme (i.e., concavely increases as $\varepsilon^{\rm{C}}$ increases) while does not strongly affect that in the TS-based SWIPT scheme (i.e., remains flat when $\varepsilon^{\rm{C}} \ge 0.012$). The reason can be explained as follows. It is intuitive to increase the SWIPT transmission power when $\varepsilon^{\rm{C}}$ increases (i.e., relaxing the constraint on detection error) so as to jointly improve the link reliability and power outage. In the PS-based SWIPT scheme, as only a small amount of the received signal power will be used for information decoding again, although the SWIPT transmission power increases along with the increase in $\varepsilon^{\rm{C}}$, a high energy efficiency in the wireless communication can be achieved and the network utility further increases. In contrast, as a large amount of the received signal power will be used for information decoding in the TS-based SWIPT scheme again, continuously increasing the SWIPT transmission power does not promise energy efficiency in wireless communication. In this case, the SWIPT transmission power stops increasing when $\varepsilon^{\rm{C}} \ge 0.012$ and thereby the network utility remains flat.

%\clearpage

\subsection{Impact of D2D Communication Distance and Adversary's Density}

\begin{figure}
		\centering
		\includegraphics[width=0.5\textwidth,trim=10 5 35 40, clip]{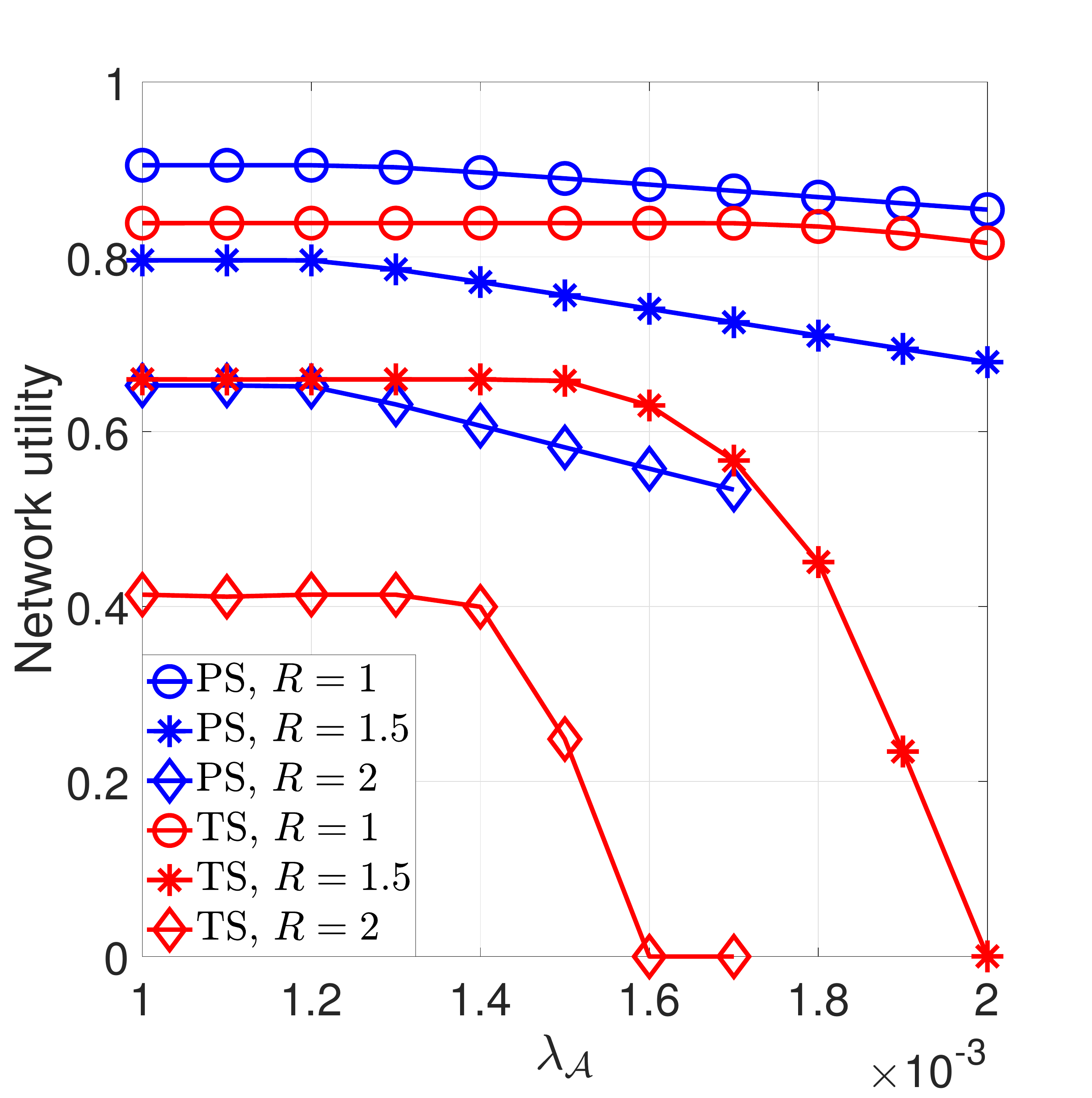}
        \caption{Impact of the D2D communication distance and adversary's density.}
        \label{fig:impact_R_lambda_A}
\end{figure}

Figure~\ref{fig:impact_R_lambda_A} evaluates the impact of the D2D communication distance (i.e., $R$) and adversary's density (i.e., $\lambda_{\cal{A}}$) on network utility. We can observe that the network utility decreases along with the increase in the adversary's density $\lambda_{\cal{A}}$. This is due to the fact that given the increase in the adversary's density $\lambda_{\cal{A}}$, the distance between the adversary and its target transmitter is likely to become shorter, which correspondingly decreases the detection error for the adversary. In this case, the transmitter has to reduce the SWIPT transmission power so as to maintain a certain level for the communication covertness, which consequently degrades its link reliability and further the network utility. Later, if the SWIPT transmission power is further reduced, the constraint on power outage will not be met (e.g., the network utilities of the PS and TS-based SWIPT schemes disappear when $R=2$ and $\lambda_{\cal{A}} > 0.0017$). Another interesting result observed from Fig.~\ref{fig:impact_R_lambda_A} is that facing more aggressive adversary (i.e., increasing adversary's density), the PS-based SWIPT scheme always outperforms the TS-based SWIPT scheme in terms of network utility especially when $R=1.5$ and $\lambda_{\cal{A}}\ge 0.0016$. This is due to the reason explained in Section~\ref{subsec:stackelberg_equilibrium} that, different from the TS-based SWIPT scheme, most of the received signal power in the PS-based SWIPT scheme will be used for energy harvesting. In this case, although the SWIPT transmission power is reduced due to the increasing threat from the adversary, the constraint on power outage in the PS-based SWIPT scheme is still easy to be met compared with that in the TS-based SWIPT scheme. Consequently, the PS-based SWIPT scheme is more robust to the adversary than the TS-based SWIPT scheme. Also, the network utility can be improved when the D2D communication distance decreases. This is due to the fact that higher network densification with lower transmission power will not only improve the communication performance but also enhance the communication covertness~\cite{9108996}. In addition, the degradation in the network utility of the PS-based SWIPT scheme (i.e., from $0.9$ to $0.65$ when $\lambda_{\cal{A}} = 0.001$) corresponding to the increase in the D2D communication distance is much smaller than that of the TS-based SWIPT scheme (i.e., from $0.84$ to $0.41$ when $\lambda_{\cal{A}} = 0.001$), which implies that the PS-based SWIPT scheme is more robust than the TS-based SWIPT scheme in terms of communication performance. In summary, the PS-based SWIPT scheme outperforms the TS-based SWIPT scheme from not only the security perspective but also from the communication efficiency perspective. 

%\clearpage

\subsection{Impact of Cellular Transmission Power and Adversary's Density}

\begin{figure}
		\centering
		\includegraphics[width=0.5\textwidth,trim=10 5 35 40, clip]{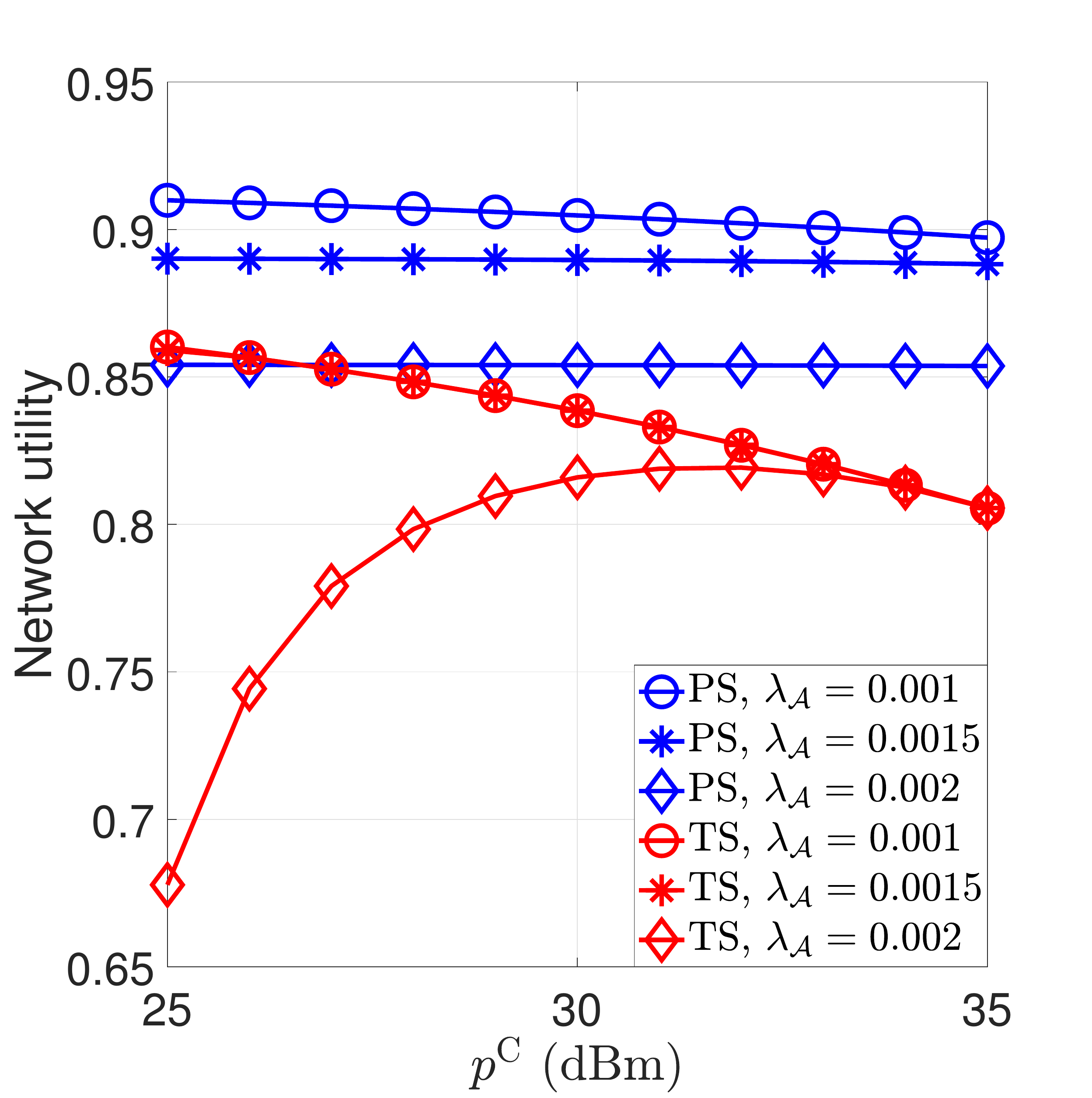}
        \caption{Impact of the cellular transmission power and adversary's density.}
        \label{fig:impact_power_C_lambda_A}
\end{figure}

Figure~\ref{fig:impact_power_C_lambda_A} evaluates the impact of the cellular transmission power $p^{\rm{C}}$ and adversary's density $\lambda_{\cal{A}}$ on the system performance. We can observe that when the adversary's density is small, the network utility decreases as $p^{\rm{C}}$ increases (i.e., $\lambda_{\cal{A}} = 0.001$ in PS-based SWIPT scheme and $\lambda_{\cal{A}} = 0.001$ and $\lambda_{\cal{A}} = 0.0015$ in TS-based SWIPT scheme). The reason is that when the adversary's density is small, it is not necessary to leverage the cellular signal to achieve covert communication on the D2D transmission. In this case, the cellular signal will damage the link reliability of the D2D transmission. In contrast, the increase in the cellular transmission power does not affect the network utility in the PS-based SWIPT scheme when the adversary's density is high (i.e., $\lambda_{\cal{A}} = 0.0015$ and $\lambda_{\cal{A}} = 0.002$ in PS-based SWIPT scheme). Moreover, the increase in the cellular transmission power first significantly improves and later degrades the network utility in TS-based SWIPT scheme when the adversary's density is high (i.e., $\lambda_{\cal{A}} = 0.002$). This can be explained as follows. The increasing cellular transmission power will be leveraged to improve the communication covertness when the adversary's density is high (i.e., more threatening adversary). As such, the D2D transmission power can be increased in the meantime, which induces a flat/increasing network utility. Regarding the decrease in the network utility of TS-based SWIPT scheme when $\lambda_{\cal{A}} = 0.002$, it is due to the fact that a stronger interference not only can improve the communication covertness but also can tamper the link reliability for the D2D transmission. This introduces a trade-off in leveraging the cellular signal to improve the communication covertness.

%\clearpage

\subsection{Impact of Threshold of Harvested Power and Cellular Transmission Power}

\begin{figure}
		\centering
		\includegraphics[width=0.5\textwidth,trim=10 5 35 40, clip]{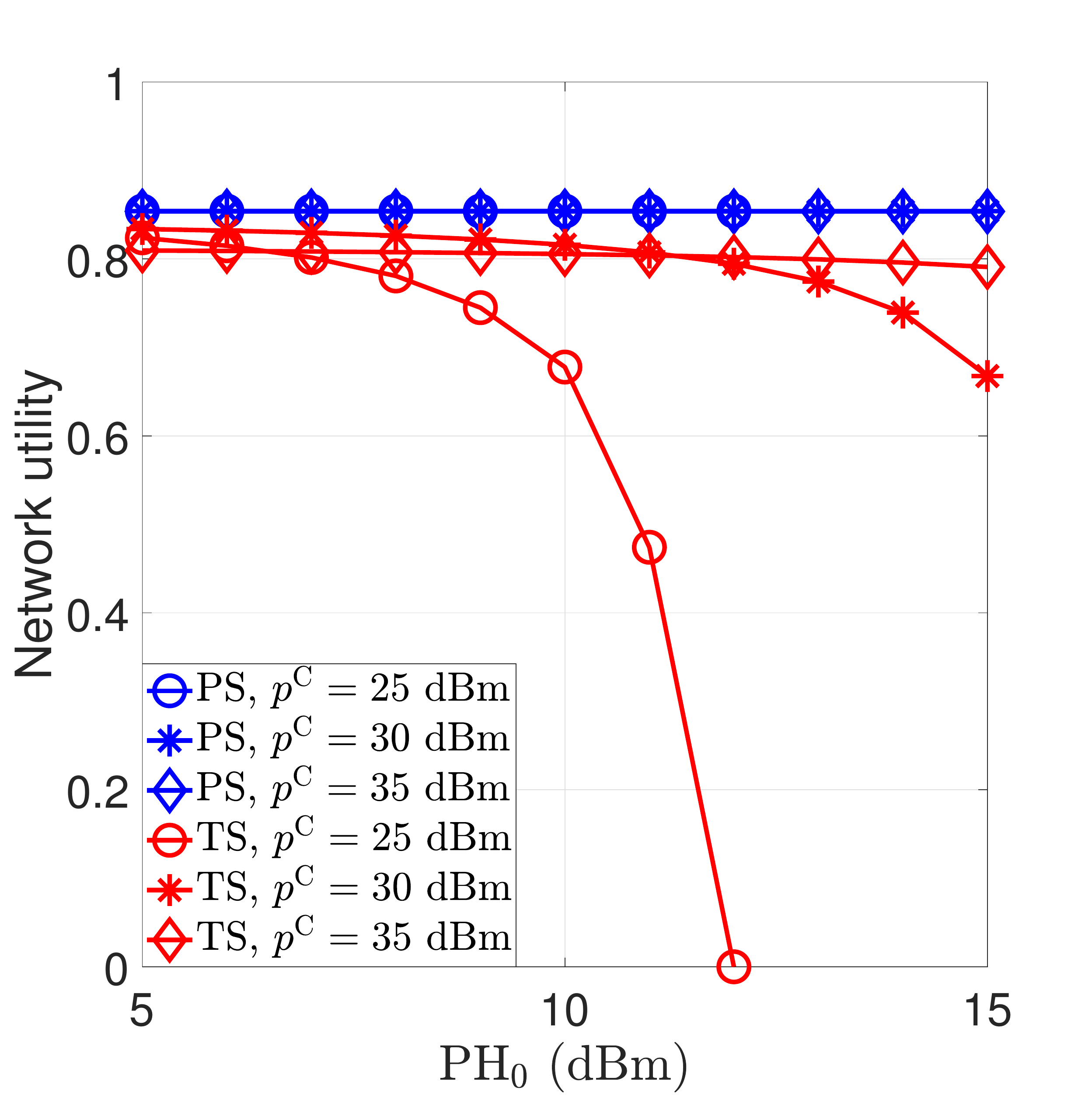}
        \caption{Impact of the threshold of harvested power and cellular transmission power.}
        \label{fig:impact_PH0_power_C}
\end{figure}

We evaluate the impact of the threshold of harvested power ${\rm{PH}}_0$ and cellular transmission power $p^{\rm{C}}$ on the network utility.  An interesting result can be observed in Fig.~\ref{fig:impact_PH0_power_C} that neither the threshold of harvested power ${\rm{PH}}_0$ nor the cellular transmission power $p^{\rm{C}}$ affects the network utility in the PS-based SWIPT scheme. In contrast, in the TS-based SWIPT scheme, the network utility with low cellular transmission power (i.e., $p^{\rm{C}} = 25$\,dBm and $p^{\rm{C}} = 30$\,dBm) is higher than that with high cellular transmission power (i.e., $p^{\rm{C}} = 35$\,dBm) when the threshold of harvested power ${\rm{PH}}_0$ is small, i.e., ${\rm{PH}}_0 \le 6$\,dBm. Later, when ${\rm{PH}}_0 \ge 12$\,dBm, the network utility with high cellular transmission power (i.e., $p^{\rm{C}} = 35$\,dBm) is higher than that with low cellular transmission power (i.e., $p^{\rm{C}} = 25$\,dBm and $p^{\rm{C}} = 30$\,dBm). Moreover, the network utility disappears when $p^{\rm{C}} = 25$\,dBm and ${\rm{PH}}_0 \ge 12$\,dBm. This is due to the fact that the constraint on power outage becomes tighter when ${\rm{PH}}_0$ increases, which will gradually degrade the network utility until this constraint cannot be satisfied. Such a result is more pronounced when the cellular transmission power is low due to the fact that it becomes harder to meet the constraint on power outage. This means that in the TS-based SWIPT scheme with a high threshold of harvested power, a higher cellular transmission power can improve network utility.

%%\clearpage

\section{Conclusion}
\label{sec:conclusion}

We have investigated a large-scale SWIPT-enabled D2D network, where covert communication has been applied to secure D2D transmissions against the adversaries from transmission detection. The D2D network underlays a downlink cellular network. The objective is to maximize the network utility subject to the constraints on detection error and power outage. The combat between the D2D network and the adversaries has been modeled by a Stackelberg game with the adversaries as the followers at the lower stage and the D2D network as the leader at the upper stage. The network spatial configuration has been modeled by using stochastic geometry so as to conduct the study from system-level perspective. For the adversaries at the lower stage, we have analyzed the problem and solved it using the Rosenbrock method.  To solve the problem of the D2D network at the upper stage, we have used the best response from the lower stage and designed a bi-level algorithm based on GA. We have presented comprehensive numerical results to evaluate the system performance and validated the optimality of the Stackelberg equilibrium. Moreover, practical insights have been revealed and summarized. In our future work, we will investigate on developing low-complexity online power control solutions to achieve covert D2D communication in large-scale SWIPT-enabled D2D networks and benchmark them against optimal solutions (e.g., Stackelberg solution).
%For future work, we will study the combat between the D2D network and adversaries on a long-term manner. 

\begin{appendices}

\section{SINR Distribution for D2D Network in PS-based SWIPT Scheme}
\label{app:PS_SINR}

Based on~(\ref{eq:Prob_SINR_PS_def}) and conditioned on ${\cal{H}}_1$, the SINR distribution at receiver $d^{\rm{Rx}}$ in the PS-based SWIPT scheme is
\begin{equation}\label{eq:Prob_SINR_PS_deriv}
\begin{aligned}
& {\mathbb{P}} \left[\left.{\rm{SINR}}^{\rm{PS}}_{d^{\rm{Rx}}} \ge 2^{\frac{l^{\rm{P}}}{\Delta t}} - 1 \right|{\cal{H}}_1\right] = {\mathbb{P}} \left[ \frac{\rho p^{\rm{S}} g_{d^{\rm{Tx}}d^{\rm{Rx}}} R^{-\alpha}}{\rho \left(I^{\rm{S}}_{d^{\rm{Rx}}} + I^{\rm{C}}_{d^{\rm{Rx}}} + N^{\rm{RF}}\right) + N_{d^{\rm{Rx}}}} \ge 2^{\frac{l^{\rm{P}}}{\Delta t}} - 1 \right]\\
= & {\mathbb{P}} \left[ g_{d^{\rm{Tx}}d^{\rm{Rx}}} \ge \frac{\left(2^{\frac{l^{\rm{P}}}{\Delta t}} - 1\right)R^\alpha}{\rho p^{\rm{S}}} \left[\rho \left(I^{\rm{S}}_{d^{\rm{Rx}}} + I^{\rm{C}}_{d^{\rm{Rx}}} + N^{\rm{RF}}\right) + N_{d^{\rm{Rx}}}\right] \right] \\
\mathop = \limits^{(a)} & {\mathbb{E}} \left[ \sum_{m = 0}^{M - 1} \frac{1}{m!}\left\{\frac{\left(2^{\frac{l^{\rm{P}}}{\Delta t}} - 1\right)R^\alpha}{\rho p^{\rm{S}}} \left[\rho \left(I^{\rm{S}}_{d^{\rm{Rx}}} + I^{\rm{C}}_{d^{\rm{Rx}}} + N^{\rm{RF}}\right) + N_{d^{\rm{Rx}}}\right]\right\}^m \right. \\ 
& \quad \times \left.\exp\left(-\frac{\left(2^{\frac{l^{\rm{P}}}{\Delta t}} - 1\right)R^\alpha}{\rho p^{\rm{S}}} \left[\rho \left(I^{\rm{S}}_{d^{\rm{Rx}}} + I^{\rm{C}}_{d^{\rm{Rx}}} + N^{\rm{RF}}\right) + N_{d^{\rm{Rx}}}\right]\right)\right] \\
\mathop = \limits^{(b)} & {\mathbb{E}} \left[ \sum_{m = 0}^{M - 1} \frac{1}{m!} \left[\frac{\left(2^{\frac{l^{\rm{P}}}{\Delta t}} - 1\right)R^\alpha}{\rho p^{\rm{S}}}\right]^m \left\{ \sum_{n = 0}^m \binom{m}{n} \left[\rho \left(I^{\rm{S}}_{d^{\rm{Rx}}} + I^{\rm{C}}_{d^{\rm{Rx}}}\right)\right]^n \left(\rho N^{\rm{RF}} + N_{d^{\rm{Rx}}}\right)^{m - n} \right\} \right. \\ 
& \quad \times \left.\exp\left(-\frac{\left(2^{\frac{l^{\rm{P}}}{\Delta t}} - 1\right)R^\alpha}{p^{\rm{S}}} \left(I^{\rm{S}}_{d^{\rm{Rx}}} + I^{\rm{C}}_{d^{\rm{Rx}}}\right)\right) \exp\left(-\frac{\left(2^{\frac{l^{\rm{P}}}{\Delta t}} - 1\right)R^\alpha}{\rho p^{\rm{S}}} \left(\rho N^{\rm{RF}} + N_{d^{\rm{Rx}}}\right)\right)\right] \\
= & \exp\left(-\frac{\left(2^{\frac{l^{\rm{P}}}{\Delta t}} - 1\right)R^\alpha}{\rho p^{\rm{S}}} \left(\rho N^{\rm{RF}} + N_{d^{\rm{Rx}}}\right)\right) \sum_{m = 0}^{M - 1} \frac{1}{m!} \left[\frac{\left(2^{\frac{l^{\rm{P}}}{\Delta t}} - 1\right)R^\alpha}{\rho p^{\rm{S}}} \right]^m \times \\ 
& \left\{ \sum_{n = 0}^m \binom{m}{n} \left(\rho N^{\rm{RF}} + N_{d^{\rm{Rx}}}\right)^{m - n} \rho^n {\mathbb{E}}_{I^{\rm{S}}_{d^{\rm{Rx}}} + I^{\rm{C}}_{d^{\rm{Rx}}}} \left[ \left(I^{\rm{S}}_{d^{\rm{Rx}}} + I^{\rm{C}}_{d^{\rm{Rx}}}\right)^n \exp\left(-\frac{\left(2^{\frac{l^{\rm{P}}}{\Delta t}} - 1\right)R^\alpha}{p^{\rm{S}}} \left(I^{\rm{S}}_{d^{\rm{Rx}}} + I^{\rm{C}}_{d^{\rm{Rx}}}\right)\right) \right]\right\},
\end{aligned}
\end{equation}
where (a) follows the gamma distribution of $g_{d^{\rm{Tx}}d^{\rm{Rx}}}$, (b) follows the binomial theorem~\cite{doi:10.1080/00029890.1949.11999350}, and
\begin{equation}\label{eq:laplace_I_dRx_derivative}
\begin{aligned}
& {\mathbb{E}}_{I^{\rm{S}}_{d^{\rm{Rx}}} + I^{\rm{C}}_{d^{\rm{Rx}}}} \left[ \left(I^{\rm{S}}_{d^{\rm{Rx}}} + I^{\rm{C}}_{d^{\rm{Rx}}}\right)^n \exp\left(-\frac{\left(2^{\frac{l^{\rm{P}}}{\Delta t}} - 1\right)R^\alpha}{p^{\rm{S}}} \left(I^{\rm{S}}_{d^{\rm{Rx}}} + I^{\rm{C}}_{d^{\rm{Rx}}}\right)\right) \right] \\
= & \left( -1 \right)^n \left\{\frac{{\rm{d}}^n}{{\rm{d}}s^n} {\mathbb{E}}_{I^{\rm{S}}_{d^{\rm{Rx}}} + I^{\rm{C}}_{d^{\rm{Rx}}}} \left[ \exp\left(- s \left(I^{\rm{S}}_{d^{\rm{Rx}}} + I^{\rm{C}}_{d^{\rm{Rx}}}\right)\right)\right]\right\}_{s = \frac{\left(2^{\frac{l^{\rm{P}}}{\Delta t}} - 1\right)R^\alpha}{p^{\rm{S}}}}.
\end{aligned}
\end{equation}
Therein,
\begin{equation}
{\mathbb{E}}_{I^{\rm{S}}_{d^{\rm{Rx}}} + I^{\rm{C}}_{d^{\rm{Rx}}}} \left[ \exp\left(- s \left(I^{\rm{S}}_{d^{\rm{Rx}}} + I^{\rm{C}}_{d^{\rm{Rx}}}\right)\right)\right] = {\mathbb{E}}_{I^{\rm{S}}_{d^{\rm{Rx}}}} \left[ \exp\left(- s I^{\rm{S}}_{d^{\rm{Rx}}}\right)\right] {\mathbb{E}}_{I^{\rm{C}}_{d^{\rm{Rx}}}} \left[ \exp\left(- s I^{\rm{C}}_{d^{\rm{Rx}}}\right)\right]
\end{equation}
due to the independence between $I^{\rm{S}}_{d^{\rm{Rx}}}$ and $I^{\rm{C}}_{d^{\rm{Rx}}}$, where
\begin{equation}\label{eq:laplace_I_dRx}
\begin{aligned}
& {\mathbb{E}}_{I^{\rm{S}}_{d^{\rm{Rx}}}} \left[ \exp\left(- s I^{\rm{S}}_{d^{\rm{Rx}}}\right)\right] \\
= & {\mathbb{E}}_{\Phi_{\left\{\left.{\cal{D}}^{\rm{Tx}}\backslash \left\{d^{\rm{Tx}}\right\}\right| d^{\rm{Rx}} \right\}}} \left[ \prod\limits_{{d^{\rm{Tx}}}' \in \left\{\left.{\cal{D}}^{\rm{Tx}}\backslash \left\{d^{\rm{Tx}}\right\}\right| d^{\rm{Rx}} \right\}} {\mathbb{E}}_{g_{{d^{\rm{Tx}}}'d^{\rm{Rx}}}} \left[\exp\left(- s p^{\rm{S}} {\mathbbm{1}}_{{d^{\rm{Tx}}}'} g_{{d^{\rm{Tx}}}'d^{\rm{Rx}}} \ell\left({\bf{x}}_{{d^{\rm{Tx}}}'}, {\bf{x}}_{d^{\rm{Rx}}}\right) \right)\right]\right]\\
= & {\mathbb{E}}_{\Phi_{\left\{\left.{\cal{D}}^{\rm{Tx}}\backslash \left\{d^{\rm{Tx}}\right\}\right| d^{\rm{Rx}} \right\}}} \left[ \prod\limits_{{d^{\rm{Tx}}}' \in \left\{\left.{\cal{D}}^{\rm{Tx}}\backslash \left\{d^{\rm{Tx}}\right\}\right| d^{\rm{Rx}} \right\}} \left(\frac{{\mathbb{P}}^{{\cal{H}}_1} }{1 + s p^{\rm{S}} \ell\left({\bf{x}}_{{d^{\rm{Tx}}}'}, {\bf{x}}_{d^{\rm{Rx}}}\right)} + {\mathbb{P}}^{{\cal{H}}_0}\right)\right]\\
\mathop = \limits^{(a)} & \exp\left(-2 \pi \lambda_{{\cal{D}}^{\rm{Tx}}} \int^\infty_0 \left[1 - \left(\frac{{\mathbb{P}}^{{\cal{H}}_1} }{1 + s p^{\rm{S}} r_{{d^{\rm{Tx}}}' d^{\rm{Rx}}}^{-\alpha}} + {\mathbb{P}}^{{\cal{H}}_0}\right)\right] r_{{d^{\rm{Tx}}}' d^{\rm{Rx}}} {\rm{d}} r_{{d^{\rm{Tx}}}' d^{\rm{Rx}}}\right)\\
= & \exp\left(-2 \pi \lambda_{{\cal{D}}^{\rm{Tx}}} {\mathbb{P}}^{{\cal{H}}_1} \int^\infty_0 \frac{s p^{\rm{S}} r_{{d^{\rm{Tx}}}' d^{\rm{Rx}}}^{-\alpha + 1}}{1 + s p^{\rm{S}} r_{{d^{\rm{Tx}}}' d^{\rm{Rx}}}^{-\alpha}} {\rm{d}} r_{{d^{\rm{Tx}}}' d^{\rm{Rx}}}\right)\\
= & \exp\left(-2 \pi \lambda_{{\cal{D}}^{\rm{Tx}}} {\mathbb{P}}^{{\cal{H}}_1} \left(s p^{\rm{S}}\right)^{\frac{2}{\alpha}}\int^\infty_0 \frac{r}{1 + r^{\alpha}} {\rm{d}} r\right)\\
= & \exp\left(-2 \pi \lambda_{{\cal{D}}^{\rm{Tx}}} {\mathbb{P}}^{{\cal{H}}_1} \left(s p^{\rm{S}}\right)^{\frac{2}{\alpha}}\int^\infty_0 r \int^\infty_0 \exp\left(-t\left(1 + r^{\alpha}\right)\right) {\rm{d}} t {\rm{d}} r\right)\\
= & \exp\left(-2 \pi \lambda_{{\cal{D}}^{\rm{Tx}}} {\mathbb{P}}^{{\cal{H}}_1} \left(s p^{\rm{S}}\right)^{\frac{2}{\alpha}}\int^\infty_0 \exp\left(-t\right) \int^\infty_0 r\exp\left(-t r^{\alpha}\right) {\rm{d}} r {\rm{d}} t\right)\\
= & \exp\left(-\pi \lambda_{{\cal{D}}^{\rm{Tx}}} {\mathbb{P}}^{{\cal{H}}_1} \left(s p^{\rm{S}}\right)^{\frac{2}{\alpha}}\int^\infty_0 \exp\left(-t\right) t^{-\frac{2}{\alpha}} {\rm{d}} t \int^\infty_0 \theta^{\frac{2}{\alpha}}\exp\left(-\theta\right) {\rm{d}} \theta \right)\\
\mathop = \limits^{(b)} & \exp\left(-\frac{\pi \lambda_{{\cal{D}}^{\rm{Tx}}} {\mathbb{P}}^{{\cal{H}}_1} \left(s p^{\rm{S}}\right)^{\frac{2}{\alpha}}}{{\rm{sinc}}\left(\frac{2}{\alpha}\right)}\right),
\end{aligned}
\end{equation}
with (a) following the probability generating functionals of PPP~\cite{9516701} and (b) following the reflection formula in Pi function\footnote{$\Pi\left(z\right) \triangleq \int^\infty_0 t^z \exp\left(-t\right) {\rm{d}} t$ is the Pi function and $\Pi\left(z\right)\Pi\left(-z\right) = \frac{\pi z}{\sin\left(\pi z\right)} = \frac{1}{{\rm{sinc}}\left(z\right)}$ is the reflection formula~\cite{9736993}.}, and ${\mathbb{E}}_{I^{\rm{C}}_{d^{\rm{Rx}}}} \left[ \exp\left(- s I^{\rm{C}}_{d^{\rm{Rx}}}\right)\right] \mathop = \limits^{(\ref{eq:laplace_I_dRx})} \exp\left(-\frac{\pi \lambda_{{\cal{B}}} {\mathbb{P}}^{{\cal{C}}_1} \left(s p^{\rm{C}}\right)^{\frac{2}{\alpha}}}{{\rm{sinc}}\left(\frac{2}{\alpha}\right)}\right)$.

\section{Distribution of Harvested Power for D2D Network in PS-based SWIPT Scheme}
\label{app:PS_energy}

Based on~(\ref{eq:Prob_PH_PS_def}) and conditioned on ${\cal{H}}_1$, the distribution of harvested power at receiver $d^{\rm{Rx}}$ in the PS-based SWIPT scheme is
\begin{equation}
\begin{aligned}
& {\mathbb{P}} \left[\left.{\rm{PH}}^{\rm{PS}}_{d^{\rm{Rx}}} \ge {\rm{PH}}_0 \right|{\cal{H}}_1\right] = {\mathbb{P}} \left[ \left(1 - \rho\right) \left[p^{\rm{S}} g_{d^{\rm{Tx}}d^{\rm{Rx}}} R^{-\alpha} + I^{\rm{S}}_{d^{\rm{Rx}}} + I^{\rm{C}}_{d^{\rm{Rx}}} + N^{\rm{RF}}\right] + N_{d^{\rm{Rx}}} \ge {\rm{PH}}_0 \right]\\
= & {\mathbb{P}} \left[ p^{\rm{S}} g_{d^{\rm{Tx}}d^{\rm{Rx}}} R^{-\alpha} + I^{\rm{S}}_{d^{\rm{Rx}}} + I^{\rm{C}}_{d^{\rm{Rx}}} \ge \frac{{\rm{PH}}_0 - N_{d^{\rm{Rx}}}}{1 - \rho} - N^{\rm{RF}}\right] \\
= & 1 - \int_0^{\frac{{\rm{PH}}_0 - N_{d^{\rm{Rx}}}}{1 - \rho} - N^{\rm{RF}}} f_{p^{\rm{S}} g_{d^{\rm{Tx}}d^{\rm{Rx}}} R^{-\alpha}}\left(t\right) F_{I^{\rm{S}}_{d^{\rm{Rx}}} + I^{\rm{C}}_{d^{\rm{Rx}}}}\left( \frac{{\rm{PH}}_0 - N_{d^{\rm{Rx}}}}{1 - \rho} - N^{\rm{RF}} - t\right) {\rm{d}} t,
\end{aligned}
\end{equation}
where 
\begin{equation}\label{eq:PDF_pg_dTxdRxR}
\begin{aligned}
f_{p^{\rm{S}} g_{d^{\rm{Tx}}d^{\rm{Rx}}} R^{-\alpha}}\left(t\right) = & \frac{{\rm{d}}}{{\rm{d}}t} F_{p^{\rm{S}} g_{d^{\rm{Tx}}d^{\rm{Rx}}} R^{-\alpha}}\left(t\right) = \frac{{\rm{d}}}{{\rm{d}}t} {\mathbb{P}}\left[p^{\rm{S}} g_{d^{\rm{Tx}}d^{\rm{Rx}}} R^{-\alpha} \le t \right] = \frac{{\rm{d}}}{{\rm{d}}t} {\mathbb{P}}\left[g_{d^{\rm{Tx}}d^{\rm{Rx}}} \le \frac{t R^{\alpha}}{p^{\rm{S}}} \right]\\
= & \frac{{\rm{d}}}{{\rm{d}}t} \frac{\gamma\left(M, \frac{t R^{\alpha}}{p^{\rm{S}}}\right)}{\Gamma\left(M\right)} = \frac{1}{\Gamma\left(M\right)}\frac{{\rm{d}}}{{\rm{d}}t} \int^{\frac{t R^{\alpha}}{p^{\rm{S}}}}_0 \theta^{M - 1} \exp\left(-\theta\right) {\rm{d}} \theta\\
\mathop = \limits^{(a)} & \frac{1}{\Gamma\left(M\right)} \frac{R^{\alpha}}{p^{\rm{S}}} \left(\frac{t R^{\alpha}}{p^{\rm{S}}}\right)^{M - 1} \exp\left(-\frac{t R^{\alpha}}{p^{\rm{S}}}\right) 
\end{aligned}
\end{equation}
with $\gamma\left(\cdot, \cdot\right)$ and $\Gamma\left(\cdot\right)$ being lower incomplete gamma function~\cite{milgram1985generalized} and gamma function~\cite{doi:10.1080/00029890.1959.11989422}, respectively, and (a) following the Leibniz integral rule~\cite{doi:10.1080/00029890.1973.11993339}, and 
\begin{equation}\label{eq:CDF_I_dRx}
\begin{aligned}
& F_{I^{\rm{S}}_{d^{\rm{Rx}}} + I^{\rm{C}}_{d^{\rm{Rx}}}}\left(t\right) \mathop = \limits^{(\ref{eq:laplace_I_a})-(\ref{eq:inverse_laplace_laplace_I_a})} {\cal{L}}^{-1}_{{\cal{L}}_{I^{\rm{S}}_{d^{\rm{Rx}}} + I^{\rm{C}}_{d^{\rm{Rx}}}}\left(s\right)}\left(t\right) \\
= & {\cal{L}}^{-1}_{{\mathbb{E}}_{I^{\rm{S}}_{d^{\rm{Rx}}} + I^{\rm{C}}_{d^{\rm{Rx}}}} \left[ \exp\left(- s \left(I^{\rm{S}}_{d^{\rm{Rx}}} + I^{\rm{C}}_{d^{\rm{Rx}}}\right)\right)\right]}\left(t\right) \mathop = \limits^{(\ref{eq:laplace_I_dRx})} {\cal{L}}^{-1}_{\exp\left(- s^{\frac{2}{\alpha}} \nu_{I^{\rm{S}}_{d^{\rm{Rx}}} + I^{\rm{C}}_{d^{\rm{Rx}}}} \right)}\left(t\right)\\
\mathop = \limits^{{\text{\cite{10.5555/1534700}}}} & 1 - \frac{1}{\pi} \int^\infty_0 \exp\left(-\nu_{I^{\rm{S}}_{d^{\rm{Rx}}} + I^{\rm{C}}_{d^{\rm{Rx}}}} \theta^{\frac{2}{\alpha}}\cos\left(\frac{2\pi}{\alpha}\right) - t\theta\right)\sin\left(\nu_{I^{\rm{S}}_{d^{\rm{Rx}}} + I^{\rm{C}}_{d^{\rm{Rx}}}} \theta^{\frac{2}{\alpha}} \sin\left(\frac{2\pi}{\alpha}\right)\right) \frac{{\rm{d}} \theta}{\theta}
\end{aligned}
\end{equation}
with $\nu_{I^{\rm{S}}_{d^{\rm{Rx}}} + I^{\rm{C}}_{d^{\rm{Rx}}}} = \frac{\pi}{{\rm{sinc}}\left(\frac{2}{\alpha}\right)}\left[\lambda_{{\cal{D}}^{\rm{Tx}}} {\mathbb{P}}^{{\cal{H}}_1} \left(p^{\rm{S}}\right)^{\frac{2}{\alpha}} + \lambda_{{\cal{B}}} {\mathbb{P}}^{{\cal{C}}_1} \left(p^{\rm{C}}\right)^{\frac{2}{\alpha}}\right]$.

\section{SINR Distribution for D2D Network in TS-based SWIPT Scheme}
\label{app:TS_SINR}

Based on~(\ref{eq:Prob_SINR_TS_def}) and conditioned on ${\cal{H}}_1$, the SINR distribution at receiver $d^{\rm{Rx}}$ in the TS-based SWIPT scheme is
\begin{equation}
\begin{aligned}
& {\mathbb{P}} \left[\left.{\rm{SINR}}^{\rm{TS}}_{d^{\rm{Rx}}} \ge 2^{\frac{l^{\rm{P}}}{\rho \Delta t}} - 1 \right|{\cal{H}}_1\right] = {\mathbb{P}} \left[ \frac{p^{\rm{S}} g_{d^{\rm{Tx}}d^{\rm{Rx}}} R^{-\alpha}}{I^{\rm{S}}_{d^{\rm{Rx}}} + I^{\rm{C}}_{d^{\rm{Rx}}} + N^{\rm{RF}} + N_{d^{\rm{Rx}}}} \ge 2^{\frac{l^{\rm{P}}}{\rho \Delta t}} - 1 \right]\\
\mathop = \limits^{(\ref{eq:Prob_SINR_PS_deriv})} & \exp\left(-\frac{\left(2^{\frac{l^{\rm{P}}}{\rho \Delta t}} - 1\right)R^\alpha}{p^{\rm{S}}} \left(N^{\rm{RF}} + N_{d^{\rm{Rx}}}\right)\right) \sum_{m = 1}^{M - 1} \frac{1}{m!} \left[\frac{\left(2^{\frac{l^{\rm{P}}}{\rho \Delta t}} - 1\right)R^\alpha}{p^{\rm{S}}}\right]^m \\
& \times \left\{ \sum_{n = 0}^{m} \binom{m}{n} \left( N^{\rm{RF}} + N_{d^{\rm{Rx}}}\right)^{m - n} {\mathbb{E}}_{I^{\rm{S}}_{d^{\rm{Rx}}} + I^{\rm{C}}_{d^{\rm{Rx}}}} \left[ \left(I^{\rm{S}}_{d^{\rm{Rx}}} + I^{\rm{C}}_{d^{\rm{Rx}}}\right)^n \exp\left(-\frac{\left(2^{\frac{l^{\rm{P}}}{\rho \Delta t}} - 1\right)R^\alpha}{p^{\rm{S}}} \left(I^{\rm{S}}_{d^{\rm{Rx}}} + I^{\rm{C}}_{d^{\rm{Rx}}}\right) \right) \right] \right\},
\end{aligned}
\end{equation}
where
\begin{equation}
\begin{aligned}
& {\mathbb{E}}_{I^{\rm{S}}_{d^{\rm{Rx}}} + I^{\rm{C}}_{d^{\rm{Rx}}}} \left[ \left(I^{\rm{S}}_{d^{\rm{Rx}}} + I^{\rm{C}}_{d^{\rm{Rx}}}\right)^n \exp\left(-\frac{\left(2^{\frac{l^{\rm{P}}}{\rho \Delta t}} - 1\right)R^\alpha}{p^{\rm{S}}} \left(I^{\rm{S}}_{d^{\rm{Rx}}} + I^{\rm{C}}_{d^{\rm{Rx}}}\right) \right) \right] \\
\mathop = \limits^{(\ref{eq:laplace_I_dRx_derivative})} & \left(-1\right)^n \left\{ \frac{{\rm{d}}^n}{{\rm{d}} s^n} {\mathbb{E}}_{I^{\rm{S}}_{d^{\rm{Rx}}} + I^{\rm{C}}_{d^{\rm{Rx}}}} \left[\exp\left(-s \left(I^{\rm{S}}_{d^{\rm{Rx}}} + I^{\rm{C}}_{d^{\rm{Rx}}}\right) \right)\right] \right\}_{s = \frac{\left(2^{\frac{l^{\rm{P}}}{\rho \Delta t}} - 1\right)R^\alpha}{p^{\rm{S}}}}
\end{aligned}
\end{equation}
with $ {\mathbb{E}}_{I^{\rm{S}}_{d^{\rm{Rx}}} + I^{\rm{C}}_{d^{\rm{Rx}}}} \left[\exp\left(-s \left(I^{\rm{S}}_{d^{\rm{Rx}}} + I^{\rm{C}}_{d^{\rm{Rx}}}\right) \right)\right]$ given in~(\ref{eq:laplace_I_dRx}).

\section{Distribution of Harvested Power for D2D Network in TS-based SWIPT Scheme}
\label{app:TS_energy}

Based on~(\ref{eq:Prob_PH_TS_def}) and conditioned on ${\cal{H}}_1$, the distribution of harvested power at receiver $d^{\rm{Rx}}$ in the TS-based SWIPT scheme is
\begin{equation}
\begin{aligned}
& {\mathbb{P}} \left[\left.{\rm{PH}}^{\rm{TS}}_{d^{\rm{Rx}}} \ge \frac{{\rm{PH}}_0}{1 - \rho}\right|{\cal{H}}_1\right] = {\mathbb{P}} \left[ p^{\rm{S}} g_{d^{\rm{Tx}}d^{\rm{Rx}}} R^{-\alpha} + I^{\rm{S}}_{d^{\rm{Rx}}} + I^{\rm{C}}_{d^{\rm{Rx}}} + N^{\rm{RF}} + N_{d^{\rm{Rx}}} \ge \frac{{\rm{PH}}_0}{1 - \rho} \right]\\
= & {\mathbb{P}} \left[ p^{\rm{S}} g_{d^{\rm{Tx}}d^{\rm{Rx}}} R^{-\alpha} + I^{\rm{S}}_{d^{\rm{Rx}}} + I^{\rm{C}}_{d^{\rm{Rx}}} \ge \frac{{\rm{PH}}_0}{1 - \rho} - N_{d^{\rm{Rx}}} - N^{\rm{RF}}\right] \\
= & 1 - \int_0^{\frac{{\rm{PH}}_0}{1 - \rho} - N_{d^{\rm{Rx}}} - N^{\rm{RF}}} f_{p^{\rm{S}} g_{d^{\rm{Tx}}d^{\rm{Rx}}} R^{-\alpha}}\left(t\right) F_{I^{\rm{S}}_{d^{\rm{Rx}}} + I^{\rm{C}}_{d^{\rm{Rx}}}}\left( \frac{{\rm{PH}}_0}{1 - \rho} - N_{d^{\rm{Rx}}} - N^{\rm{RF}} - t \right) {\rm{d}} t,
\end{aligned}
\end{equation}
where $f_{p^{\rm{S}} g_{d^{\rm{Tx}}d^{\rm{Rx}}} R^{-\alpha}}\left(\cdot\right)$ and $F_{I^{\rm{S}}_{d^{\rm{Rx}}} + I^{\rm{C}}_{d^{\rm{Rx}}}}\left(\cdot\right)$ are given in~(\ref{eq:PDF_pg_dTxdRxR}) and~(\ref{eq:CDF_I_dRx}), respectively.

\section{False Alarm Probability For Adversary}
\label{app:FA_prob}

According to~(\ref{eq:FA_a}) and conditioned on ${\cal{H}}_0$, the FA probability of adversary $a$ is 
\begin{equation}
P^{\rm{FA}}_a\left(p^{\rm{S}}, \tau\right) = {\mathbb{P}}\left[\left.y_{a} > \tau \right| {\cal{H}}_{0}\right] = {\mathbb{P}}\left[ I^{\rm{S}}_{a} + I^{\rm{C}}_{a} > \tau - N_a \right] = 1 - F_{I^{\rm{S}}_{a} + I^{\rm{C}}_{a}}\left(\tau - N_a\right),
\end{equation}
where $F_{I^{\rm{S}}_{a} + I^{\rm{C}}_{a}}\left(\cdot\right)$ can be obtained by~\cite{7056528}
\begin{enumerate}
\item Laplace transform of $I^{\rm{S}}_{a} + I^{\rm{C}}_{a}$:
 \begin{equation}\label{eq:laplace_I_a}
 {\cal{L}}_{I^{\rm{S}}_{a} + I^{\rm{C}}_{a}}\left(s\right) = {\mathbb{E}}_{I^{\rm{S}}_{a} + I^{\rm{C}}_{a}} \left[ \exp\left(- s \left(I^{\rm{S}}_{a} + I^{\rm{C}}_{a}\right)\right)\right] \mathop = \limits^{(\ref{eq:laplace_I_dRx})} \exp\left(-s^{\frac{2}{\alpha}} \nu_{I^{\rm{S}}_{a} + I^{\rm{C}}_{a}}\right)
\end{equation}

\item Inverse Laplace transform of the Laplace transform of $I^{\rm{S}}_{a} + I^{\rm{C}}_{a}$\footnote{Please refer to the Bromwich inversion theorem in Chapter~2 of~\cite{10.5555/1534700} for detailed derivation.}: 
\begin{equation}\label{eq:inverse_laplace_laplace_I_a}
\begin{aligned}
& F_{I^{\rm{S}}_{a} + I^{\rm{C}}_{a}}\left(t\right) = {\cal{L}}^{-1}_{{\cal{L}}_{I^{\rm{S}}_{a} + I^{\rm{C}}_{a}}\left(s\right)}\left(t\right) \\
= & 1 - \frac{1}{\pi} \int^\infty_0 \exp\left(-\nu_{I^{\rm{S}}_{a} + I^{\rm{C}}_{a}} \theta^{\frac{2}{\alpha}}\cos\left(\frac{2\pi}{\alpha}\right) - t\theta\right)\sin\left(\nu_{I^{\rm{S}}_{a} + I^{\rm{C}}_{a}} \theta^{\frac{2}{\alpha}} \sin\left(\frac{2\pi}{\alpha}\right)\right) \frac{{\rm{d}} \theta}{\theta}
\end{aligned}
\end{equation}
with $\nu_{I^{\rm{S}}_{a} + I^{\rm{C}}_{a}} = \frac{\pi}{{\rm{sinc}}\left(\frac{2}{\alpha}\right)}\left[\lambda_{{\cal{D}}^{\rm{Tx}}} {\mathbb{P}}^{{\cal{H}}_1} \left(p^{\rm{S}}\right)^{\frac{2}{\alpha}} + \lambda_{{\cal{B}}} {\mathbb{P}}^{{\cal{C}}_1} \left(p^{\rm{C}}\right)^{\frac{2}{\alpha}}\right]$.
\end{enumerate}

\section{Miss Detection Probability For Adversary}
\label{app:MD_prob}

According to~(\ref{eq:MD_a}) and conditioned on ${\cal{H}}_1$, the MD probability of adversary $a$ is 
\begin{equation}
\begin{aligned}
P^{\rm{MD}}_a\left(p^{\rm{S}}, \tau\right) = & {\mathbb{P}}\left[\left.y_{a} < \tau \right| {\cal{H}}_{1}\right] = {\mathbb{P}}\left[ p^{\rm{S}} g_{d^{\rm{Tx}} a} \ell\left({\bf{x}}_{d^{\rm{Tx}}}, {\bf{x}}_a\right) + I^{\rm{S}}_{a} + I^{\rm{C}}_{a} < \tau - N_a \right] \\
= & \int^{\tau - N_a}_0 f_{p^{\rm{S}} g_{d^{\rm{Tx}} a} \ell\left({\bf{x}}_{d^{\rm{Tx}}}, {\bf{x}}_a\right)}\left(t\right) F_{I^{\rm{S}}_{a} + I^{\rm{C}}_{a}} \left(\tau - N_a - t\right) {\rm{d}} t,
\end{aligned}
\end{equation}
where $F_{I^{\rm{S}}_{a} + I^{\rm{C}}_{a}} \left(\cdot\right)$ has been given in~(\ref{eq:inverse_laplace_laplace_I_a}) and
\begin{equation}\label{eq:PDF_pg_dTxa_l}
\begin{aligned}
f_{p^{\rm{S}} g_{d^{\rm{Tx}} a} \ell\left({\bf{x}}_{d^{\rm{Tx}}}, {\bf{x}}_a\right)}\left(t\right) = & \frac{{\rm{d}}}{{\rm{d}}t} F_{p^{\rm{S}} g_{d^{\rm{Tx}} a} \ell\left({\bf{x}}_{d^{\rm{Tx}}}, {\bf{x}}_a\right)}\left(t\right) = \frac{{\rm{d}}}{{\rm{d}}t} {\mathbb{P}}\left[p^{\rm{S}} g_{d^{\rm{Tx}} a} \ell\left({\bf{x}}_{d^{\rm{Tx}}}, {\bf{x}}_a\right) \le t \right] \\
= & \frac{{\rm{d}}}{{\rm{d}}t} {\mathbb{P}}\left[g_{d^{\rm{Tx}} a} \le \frac{t r_{d^{\rm{Tx}} a}^{\alpha}}{p^{\rm{S}}} \right] = \frac{{\rm{d}}}{{\rm{d}}t} {\mathbb{E}}_{r_{d^{\rm{Tx}} a}}\left[1 - \exp\left(-\frac{t r_{d^{\rm{Tx}} a}^{\alpha}}{p^{\rm{S}}}\right) \right]\\
= & \int^\infty_0 f_{r_{d^{\rm{Tx}} a}}\left(r\right) \exp\left(-\frac{t r^{\alpha}}{p^{\rm{S}}}\right) \frac{r^{\alpha}}{p^{\rm{S}}} {\rm{d}} r
\end{aligned}
\end{equation}
with
\begin{equation}
f_{r_{d^{\rm{Tx}} a}}\left(r\right) = 2 \pi \lambda_{{\cal{A}}} r \exp\left(-\pi \lambda_{{\cal{A}}} r^2\right)
\end{equation}
following the fact that no adversary is nearer to transmitter $d^{\rm{Tx}}$ than adversary $a$\footnote{Please refer to Section~III-A of~\cite{6042301} for the details.}.

\end{appendices}

\bibliography{bibfile}

\begin{thebibliography}{10}

\bibitem{feng2022securing}
S.~Feng, X.~Lu, S.~Sun, D.~Niyato  and E.~Hossain,
\newblock ``Securing large-scale d2d networks using covert communication and
  friendly jamming,''
\newblock {\em arXiv preprint arXiv:2209.15170}, 2022.

\bibitem{SWIPT2023feng}
S.~Feng, X.~Lu, K.~Zhu, D.~Niyato  and P.~Wang,
\newblock ``Covert d2d communication underlaying cellular network: A
  system-level security perspective,''
\newblock {\em arXiv preprint arXiv:2302.01745}, 2023.

\bibitem{6609136}
K.~Huang,
\newblock ``Spatial throughput of mobile ad hoc networks powered by energy
  harvesting,''
\newblock {\em IEEE Transactions on Information Theory}, vol. 59, no. 11, pp.
  7597--7612, 2013.

\bibitem{8896400}
W.~Wang, G.~Feng, B.~Li, Y.~Yuan, Q.~Li, H.~Lv  and Q.~Zhao,
\newblock ``An online computation offloading with energy-harvesting in mobile
  ad hoc network,''
\newblock in {\em 2019 IEEE International Conference on Smart Internet of
  Things (SmartIoT)}, 2019, pp. 22--27.

\bibitem{9385384}
Y.~Xu, H.~Sun  and Y.~Ye,
\newblock ``Distributed resource allocation for swipt-based cognitive ad-hoc
  networks,''
\newblock {\em IEEE Transactions on Cognitive Communications and Networking},
  vol. 7, no. 4, pp. 1320--1332, 2021.

\bibitem{9108996}
X.~Lu, E.~Hossain, T.~Shafique, S.~Feng, H.~Jiang  and D.~Niyato,
\newblock ``Intelligent reflecting surface enabled covert communications in
  wireless networks,''
\newblock {\em IEEE Network}, vol. 34, no. 5, pp. 148--155, 2020.

\bibitem{6951347}
X.~Lu, P.~Wang, D.~Niyato, D.~I. Kim  and Z.~Han,
\newblock ``Wireless networks with rf energy harvesting: A contemporary
  survey,''
\newblock {\em IEEE Communications Surveys \& Tutorials}, vol. 17, no. 2, pp.
  757--789, 2015.

\bibitem{7457656}
I.~Krikidis,
\newblock ``Swipt in 3-d bipolar ad hoc networks with sectorized antennas,''
\newblock {\em IEEE Communications Letters}, vol. 20, no. 6, pp. 1267--1270,
  2016.

\bibitem{7063588}
C.~Song, C.~Ling, J.~Park  and B.~Clerckx,
\newblock ``Mimo broadcasting for simultaneous wireless information and power
  transfer: Weighted mmse approaches,''
\newblock in {\em 2014 IEEE Globecom Workshops (GC Wkshps)}, 2014, pp.
  1151--1156.

\bibitem{7081080}
Z.~Ding, C.~Zhong, D.~Wing Kwan~Ng, M.~Peng, H.~A. Suraweera, R.~Schober  and
  H.~V. Poor,
\newblock ``Application of smart antenna technologies in simultaneous wireless
  information and power transfer,''
\newblock {\em IEEE Communications Magazine}, vol. 53, no. 4, pp. 86--93, 2015.

\bibitem{8355734}
J.~Hu, S.~Yan, X.~Zhou, F.~Shu, J.~Li  and J.~Wang,
\newblock ``Covert communication achieved by a greedy relay in wireless
  networks,''
\newblock {\em IEEE Transactions on Wireless Communications}, vol. 17, no. 7,
  pp. 4766--4779, 2018.

\bibitem{9382022}
X.~Jiang, X.~Chen, J.~Tang, N.~Zhao, X.~Y. Zhang, D.~Niyato  and K.-K. Wong,
\newblock ``Covert communication in uav-assisted air-ground networks,''
\newblock {\em IEEE Wireless Communications}, vol. 28, no. 4, pp. 190--197,
  2021.

\bibitem{9361424}
K.-W. Huang, H.~Deng  and H.-M. Wang,
\newblock ``Jamming aided covert communication with multiple receivers,''
\newblock {\em IEEE Transactions on Wireless Communications}, vol. 20, no. 7,
  pp. 4480--4494, 2021.

\bibitem{9736993}
S.~Feng, X.~Lu, S.~Sun  and D.~Niyato,
\newblock ``Mean-field artificial noise assistance and uplink power control in
  covert iot systems,''
\newblock {\em IEEE Transactions on Wireless Communications}, vol. 21, no. 9,
  pp. 7358--7373, 2022.

\bibitem{9580594}
N.~T.~T. Van, N.~C. Luong, H.~T. Nguyen, F.~Shaohan, D.~Niyato  and D.~I. Kim,
\newblock ``Latency minimization in covert communication-enabled federated
  learning network,''
\newblock {\em IEEE Transactions on Vehicular Technology}, vol. 70, no. 12, pp.
  13447--13452, 2021.

\bibitem{7805182}
B.~He, S.~Yan, X.~Zhou  and V.~K.~N. Lau,
\newblock ``On covert communication with noise uncertainty,''
\newblock {\em IEEE Communications Letters}, vol. 21, no. 4, pp. 941--944,
  2017.

\bibitem{4712724}
A.~M. Hunter, J.~G. Andrews  and S.~Weber,
\newblock ``Transmission capacity of ad hoc networks with spatial diversity,''
\newblock {\em IEEE Transactions on Wireless Communications}, vol. 7, no. 12,
  pp. 5058--5071, 2008.

\bibitem{7150338}
X.~Zhou, J.~Guo, S.~Durrani  and I.~Krikidis,
\newblock ``Performance of maximum ratio transmission in ad hoc networks with
  swipt,''
\newblock {\em IEEE Wireless Communications Letters}, vol. 4, no. 5, pp.
  529--532, 2015.

\bibitem{875282}
A.~Shah and A.~Haimovich,
\newblock ``Performance analysis of maximal ratio combining and comparison with
  optimum combining for mobile radio communications with cochannel
  interference,''
\newblock {\em IEEE Transactions on Vehicular Technology}, vol. 49, no. 4, pp.
  1454--1463, 2000.

\bibitem{7320989}
Y.~Liu, L.~Wang, S.~A. Raza~Zaidi, M.~Elkashlan  and T.~Q. Duong,
\newblock ``Secure d2d communication in large-scale cognitive cellular
  networks: A wireless power transfer model,''
\newblock {\em IEEE Transactions on Communications}, vol. 64, no. 1, pp.
  329--342, 2016.

\bibitem{6692447}
S.~A. Mousavifar and C.~Leung,
\newblock ``Trust-based energy efficient spectrum sensing in cognitive radio
  networks,''
\newblock in {\em 2013 IEEE 78th Vehicular Technology Conference (VTC Fall)},
  2013, pp. 1--6.

\bibitem{cates2019cauchy}
D.~M. Cates,
\newblock {\em Cauchy's Calcul Infinit{\'e}simal}, vol.~9,
\newblock Springer.

\bibitem{bazaraa2013nonlinear}
M.~S. Bazaraa, H.~D. Sherali  and C.~M. Shetty,
\newblock {\em Nonlinear programming: theory and algorithms},
\newblock John Wiley \& Sons, 2013.

\bibitem{OLIVETO201521}
P.~S. Oliveto and C.~Witt,
\newblock ``Improved time complexity analysis of the simple genetic
  algorithm,''
\newblock {\em Theoretical Computer Science}, vol. 605, pp. 21--41, 2015.

\bibitem{doi:10.1080/00029890.1949.11999350}
J.~L. Coolidge,
\newblock ``The story of the binomial theorem,''
\newblock {\em The American Mathematical Monthly}, vol. 56, no. 3, pp.
  147--157, 1949.

\bibitem{9516701}
X.~Lu, M.~Salehi, M.~Haenggi, E.~Hossain  and H.~Jiang,
\newblock ``Stochastic geometry analysis of spatial-temporal performance in
  wireless networks: A tutorial,''
\newblock {\em IEEE Communications Surveys \& Tutorials}, vol. 23, no. 4, pp.
  2753--2801, 2021.

\bibitem{milgram1985generalized}
M.~Milgram,
\newblock ``The generalized integro-exponential function,''
\newblock {\em Mathematics of computation}, vol. 44, no. 170, pp. 443--458,
  1985.

\bibitem{doi:10.1080/00029890.1959.11989422}
P.~J. Davis,
\newblock ``Leonhard euler's integral: A historical profile of the gamma
  function,''
\newblock {\em The American Mathematical Monthly}, vol. 66, no. 10, pp.
  849--869, 1959.

\bibitem{doi:10.1080/00029890.1973.11993339}
H.~Flanders,
\newblock ``Differentiation under the integral sign,''
\newblock {\em The American Mathematical Monthly}, vol. 80, no. 6, pp.
  615--627, 1973.

\bibitem{10.5555/1534700}
A.~M. Cohen,
\newblock {\em Numerical Methods for Laplace Transform Inversion},
\newblock Springer Publishing Company, Incorporated, 1st edition, 2007.

\bibitem{7056528}
A.~H. Sakr and E.~Hossain,
\newblock ``Cognitive and energy harvesting-based d2d communication in cellular
  networks: Stochastic geometry modeling and analysis,''
\newblock {\em IEEE Transactions on Communications}, vol. 63, no. 5, pp.
  1867--1880, 2015.

\bibitem{6042301}
J.~G. Andrews, F.~Baccelli  and R.~K. Ganti,
\newblock ``A tractable approach to coverage and rate in cellular networks,''
\newblock {\em IEEE Transactions on Communications}, vol. 59, no. 11, pp.
  3122--3134, 2011.

\end{thebibliography}

\end{document}